\documentclass[12pt,preprint]{aastex}
\usepackage{amssymb}

\newcommand{\xmm}{{\it XMM-Newton}}
\newcommand{\sax}{{\it BeppoSAX}}

\newcommand{\et}{et al.\ }

\newcommand{\fvar}{$F_{\rm var}$}
\newcommand{\ecross}{E_{\rm cross}}

\shortauthors{Zhang}

\shorttitle{XMM-Newton observation of S5~0716+714}

\begin{document}

\title{Evolution of the synchrotron and inverse Compton emission of the low energy peaked BL Lac object S5~0716+714}

\author{Y.H. Zhang}
\affil{Department of Physics and Tsinghua Center for Astrophysics (THCA), Tsinghua University, Beijing 100084, China}
\email{youhong.zhang@mail.tsinghua.edu.cn}

\begin{abstract}

This paper presents a detailed analysis of temporal and spectral
variability of the low energy peaked BL Lac object S5~0716+714 with
a long ($\sim 74$~ks) X-ray observation performed by \xmm\ on 2007
September 24--25. The source experiences recurrent flares on
timescales of hours. The soft X-ray variations, up to a factor of
$\sim 4$, are much stronger than the hard X-ray variations. With
higher energy, the variability amplitude increases in the soft
X-rays but decreases in the hard X-rays. The hard X-ray variability
amplitude, however, is effectively large. For the first time, we
detect a soft lag of $\sim1000$~s between the soft and hard X-ray
variations. The soft lags might become larger with larger energy
differences. The overall X-ray spectra exhibit a
softer-when-brighter trend, whereas the soft X-ray spectra appear to
show a harder-when-brighter trend. The concave X-ray spectra of the
source can be interpreted as the sum of the high energy tail of the
synchrotron emission, dominating in the soft X-rays, and the low
energy end of the inverse Compton (IC) emission, contributing more
in the hard X-rays. The synchrotron spectra are steep
($\Gamma\sim2.6$), while the IC spectra are flat ($\Gamma\sim1.2$).
The synchrotron spectra appear to harden with larger synchrotron
fluxes, while the IC spectra seem to soften with larger IC fluxes.
When the source brightens, the synchrotron fluxes increase but the
IC fluxes decrease. The synchrotron tail exhibits larger flux
variations but smaller spectral changes than the IC component does.
The crossing energies between the two components and the trough
energies of spectral energy distributions (SEDs) increase when the
source brightens. The X-ray spectral variability demonstrates that
the synchrotron and IC SED peaks of S5~0716+714 shift to higher
energies when it brightens. The temporal variability also elucidates
that the hard X-ray variations of the source might be dominated by
the synchrotron tail. The simultaneous optical and UV data obtained
with \xmm\ are compared with the X-ray observations.

\end{abstract}



\section{Introduction}\label{sec:intro}

Blazars, an assembly of BL Lac objects and flat spectrum radio
quasars (FSRQs), are the most extreme class of Active Galactic
Nuclei (AGN). They are remarkably characterized by variability of
different amplitude on various timescales across most of the
electromagnetic wavelengths (e.g., Ulrich et al. 1997). The
radiation from blazars is thought to originate in a relativistic jet
closely aligned with the line of sight, which is thus
relativistically beamed (e.g., Urry \& Padovani 1995). In the
$\log(\nu F_{\nu})-\log(\nu)$ representation, the spectral energy
distributions (SEDs) of blazars comprise of two broad bumps. The low
energy bump peaks at the frequencies ranging from sub-millimeter to
X-ray bands, while the peak of the high energy bump is thought to be
at the MeV-TeV gamma-ray bands though it has not been well-known for
most of blazars yet. The low energy component is widely believed to
be the synchrotron emission of a relativistic electron population
residing in the jet tangled with magnetic field, whereas the high
energy component is thought to be the Inverse Compton (IC) radiation
of the same electron population, scattering the low energy photons
of either its own synchrotron photons (the synchrotron self-Compton
[SSC, e.g., Maraschi et al. 1992] model mostly adopted for BL Lac
objects) or the external photons of surrounding environment (the
External Compton [EC, e.g., Sikora et al. 1994] model mostly used
for FSRQs).

The current classification of blazars is preferably based on the
peak energy of the synchrotron emission component. BL Lac objects
are differentiated into high energy peaked BL Lac objects (HBLs) and
low energy-peaked BL Lac objects (LBLs) (Giommi \& Padovani 1994;
Padovani \& Giommi 1995). The synchrotron emission of HBLs and LBLs
peaks at the UV/X-ray and the IR-optical wavelengths, respectively.
The synchrotron peak of FSRQs might shift down to lower (e.g.,
sub-millimeter) frequencies. The location of the synchrotron peak is
thought to be related with the bolometric luminosity of the source:
the higher the luminosity, the lower the peak energy (Fossati et al.
1998; Ghisellini et al. 1998). This is the so-called blazar SED
sequence widely cited in the blazar studies. Historically, LBLs and
HBLs are best observed and studied in the optical and X-ray bands,
respectively, because LBLs are usually the brightest and most
variable objects in the optical wavelengths, whereas HBLs are the
brightest and most variable ones in the X-rays. The optical
variability properties of the LBL BL Lacertae are analogous to the
X-ray variability properties of bright HBLs (Papadakis et al. 2003;
Bian et al. 2007). This similarity is anticipated, since the optical
emission of LBLs and the X-ray emission of HBLs correspond to the
synchrotron peak of their own, respectively. Because it originates
in a different dominance of the synchrotron over the IC radiation,
the X-ray emission of different classes of blazars shows very
different characteristics of temporal and spectral variability (see
Pian 2002 and Zhang 2003 for reviews).

HBLs are bright and best studied X-ray sources. Their X-ray emission
is commonly interpreted as the synchrotron radiation from the
high-energy tail of an electron distribution, which is sensitive to
the particle acceleration and cooling and thus shows rapid and
strong variability (e.g., Kirk et al. 1998). Repeated X-ray
observations have been performed for a few X-ray bright HBLs with
various X-ray telescopes. Although it is complicated, the temporal
and spectral variability of HBLs might be characterized as follows.
The X-ray fluxes of HBLs exhibit large amplitude variability on
different timescales. Rapid variations are common, e.g., the fluxes
of PKS~2155--304 changed by a factor of $\sim2$ on timescales of the
order of a few hours (Sembay et al. 1993; Zhang et al. 1999). The
variability amplitude increases with higher energy (Mrk~421: Ravasio
et al. 2004; PKS~2155--304: Zhang et al. 2005, 2006b; Mrk~510:
Gliozzi et al. 2006), in accordance with the fact that the higher
energy electrons have shorter cooling timescales. The X-ray spectra
of HBLs are steep ($\Gamma > 2$) and show a convex shape. More
precisely, the X-ray spectra continuously steepen with energies
(e.g., Perman et al. 2005; Tramacere et al. 2007). Flux changes are
frequently accompanied by spectral variability. The spectra
typically harden with higher fluxes (e.g., Mrk~421: Brinkmann et al.
2005; PKS~2155-304: Sembay et al. 2002), i.e., the so-called
harder-when-brighter phenomenon. The synchrotron peaks have been
determined to locate in the X-ray band for a few bright HBLs, which
shift to higher energies as the source brightens (e.g., Mrk~421:
Fossati et al. 2000; Massaro et al. 2004a; Tanihata et al. 2004;
Tramacere et al. 2009; Mrk~501: Massaro et al. 2004b). The
variations between different energies are well correlated within the
X-ray band, often with time lags. The lags, when detected, appear to
change from flare to flare (e.g., Mrk~421: Takahashi et al. 2000;
PKS~2155--304: Zhang et al. 2002; 2006a). The variations at lower
energies often lag behind those at higher energies (soft lag),
although the opposite behavior (hard lag) has been also claimed
(Mrk~421: Zhang 2002; Ravasio et al. 2004; Brinkmann et al. 2003;
PKS 2155--305: Zhang et al. 2006a; 1ES~1218+304: Sato et al. 2008).
With the \xmm\ PN timing mode X-ray observations of Mrk~421,
Brinkmann et al. (2005) claimed that both the sign and the length of
lags may change on timescales of a few thousand seconds. The lags
are also energy dependent, increasing with larger energy differences
(e.g., Mrk~421: Takahashi et al. 1996; Zhang 2002; Zhang et al.
2010; PKS~2155--304: Kataoka et al. 2000; Zhang et al. 2002).
Spectral variability may correspond to the existence of the
inter-band time lags. Spectral evolution with flux has been
systematically resolved for some well-defined flares. On the
spectral index--flux ($\alpha-F$) plot, clockwise and
counter-clockwise loops have been effectively observed to correspond
to the soft and hard lag (e.g., Mrk~421: Takahashi et al. 1996;
Zhang 2002, 2010; Ravasio et al. 2004; PKS~2155--304: Kataoka et al.
2000; Zhang et al. 2002), respectively. The sign of the lags and the
pattern of the loops have been thought to be the results of the
balance between the acceleration and cooling timescales of the
electron population responsible for the observed X-ray emission
(Kirk et al. 1998; Zhang et al. 2002). This in turn has been used to
constrain some of the physical parameters (the magnetic field and
the Doppler factor) of the emitting region.

The X-ray spectra of LBLs have a concave shape with a break energy
of $\sim2$~keV, which has been clearly detected in a few bright
objects with the synchrotron emission peaking in the optical band
(e.g., BL Lacertae: Tanihata et al. 2000; Ravasio et al. 2002, 2003;
S5~0716+714: Wagner et al. 1996; ON~231: Tagliaferri et al. 2000).
The spectra are steep ($\Gamma \sim 2.3$) in the soft X-rays, while
they are flat ($\Gamma \sim 1.7$) in the hard X-rays. On timescale
of hours, the soft X-ray fluxes are highly variable, whereas the
hard X-ray fluxes tend to be stable (e.g., Tanihata et al. 2000;
Ravasio et al. 2003; Giommi et al. 1999; Tagliaferri et al. 2000).
Therefore, LBLs shows different spectral and temporal properties in
the soft and hard X-rays, which is thought to be due to different
emission components. The soft X-rays are dominated by the
synchrotron emission from the high-energy tail of an electron
distribution, whereas the IC scattering of the low-energy side of
the same electron distribution off the synchrotron (and/or external
in some cases) photons contributes more in the hard X-rays (e.g.,
Giommi et al. 1999). The high energy tail of the synchrotron
emission is variable and has a steep spectrum. The low energy side
of the IC emission is stable and its spectrum is flat. Accordingly,
the studies of LBLs in the X-rays are able to understand the
physical conditions of the low- and high-energy electrons
simultaneously. By fitting the X-ray spectra with the double power
law model, Tanihata et al. (2000) and Ferrero et al. (2006)
disentangled the two emission components for BL Lacertae and
S5~0716+714. The synchrotron emission is variable on short
timescales (e.g., $\sim$~hours), whereas the IC emission appears to
vary on longer timescales (e.g., $\sim$~days). The breaking energies
become higher with higher fluxes (e.g., BL Lacertae: Ravasio et al.
2003). The overall X-ray spectra show the softer-when-brighter
behaviour (e.g., Giommi et al. 1999. These phenomena are ascribed to
the upshift of the synchrotron peak to higher energy when the
sources brighten. No inter-band time lags or the spectral
index--flux loops have been clearly reported so far for LBLs,
although they are theoretically expected (B${\rm \ddot{o}}$ttcher \&
Chiang 2002).

S5~0716+714 is identified as a prototype of LBLs since its
synchrotron emission peaks in the optical band (e.g., Nieppola et
al. 2006). The photometric detection of its host galaxy suggested a
redshift of $0.31\pm0.08$ (Nilsson et al. 2008). It has been
intensively observed and studied in many wavelengths, especially in
the optical band (e.g., Wu et al. 2007 and reference therein). It is
strongly variable from radio to X-ray bands on different timescales
(e.g., Wagner et al. 1996). The EGRET onboard the Compton Gamma-ray
Observatory detected its high energy gamma-ray ($>100$~MeV) emission
several times from 1991 to 1996 (Hartman et al. 1999). The gamma-ray
fluxes varied by a factor of two on timescale of years, but the
spectral index--flux correlation was not found (Nandikotkur et al.
2007). In 2008, AGILE detected a variable gamma-ray flux, with a
peak flux above the maximum obtained by EGRET (Chen et al. 2008).
Observations by HEGRA resulted in an upper limit of flux at very
high energy (VHE) gamma-ray energies ($>1.6$~TeV) (Aharonian et al.
2004). It is also in the first Fermi-LAT bright source list (Abdo et
al. 2009). Recently, the MAGIC collaboration reported the first
detection of VHE gamma-rays from the source at $5.8\sigma$ level
(Anderhub et al. 2009). The discovery of S5~0716+714 as a VHE
gamma-ray LBL was triggered by its very high optical state,
suggesting a possible correlation between the VHE gamma-ray and the
optical emissions. S5~0716+714 is also a preferred target to perform
simultaneous multi-wavelength observations (e.g., Villata et al.
2008; Giommi et al. 2008). Wu et al. (2009) reported the first
detection of time lags between the variations in different optical
wavelengths.

S5~0716+71 has been observed by various X-ray telescopes. The
observations in 1991 March with the PSPC onboard ROSAT revealed a
behavior of flux-related spectral variations in the 0.1--2.4~keV
band. Two distinct spectral components are needed to describe the
concave X-ray spectra (Cappi et al. 1994). ASCA observed the source
in 1994 March and confirmed the spectral shape found by ROSAT in
higher energies: the spectra flatten with increasing energies (Kubo
et al. 1998). The three \sax\ observations in 1996, 1998 and 2000
revealed the spectral and temporal variability of the two spectral
components. It was in faint states in the 1996 and 1998 observation
(Giommi et al. 1999). The spectral fits with a broken power law
model resulted in concave spectra in the 0.1--10~keV band, breaking
at $\sim 2-3$~keV. The 2000 observation caught the source in its
high state (Tagliaferri et al. 2003). The concave spectra can be
disentangled into two power law components, crossing at
$\sim1.5$~keV. The steeper power law component dominates the soft
X-ray emission, whereas the flatter one contributes more to the hard
X-rays. The soft X-ray spectral index is the largest out of the
three \sax\ observations, indicating a softer-when-brighter
behaviour for the soft X-ray variability. The \sax\ observations
also revealed large and rapid variations in the soft X-ray band and
the lack of variations above $\sim3$~keV. In the 1996 observation, a
flare with duration of $\sim20$~hours was detected in the soft but
not in the hard X-ray band (Giommi et al. 1999). Variations by a
factor of $\sim3$ on a timescale of one hour was detected below
$\sim3$~keV and no variations above (Tagliaferri et al. 2003).
Nevertheless, the comparisons between the three observations show
that the hard X-ray fluxes were also variable over the timescales of
years. The temporal and spectral variability of S5~0716+714 is in
good agreement with the interpretation that the soft and hard X-ray
emission, separating at $\sim2-3$~keV, are dominated by the strongly
variable high energy tail of the synchrotron radiation and the
"stable" low energy side of the IC radiation, respectively. The
softer-when-brighter behaviour, opposite to the harder-when-brighter
one frequently observed in HBLs, could be ascribed by the upshift of
the synchrotron peak to higher energy as the source brightens,
causing the synchrotron tail to enter into the soft X-ray band more
(e.g., Giommi et al. 1999).

However, previous X-ray observations were not able to disentangle
the two spectral components and related variations on short
timescales. \xmm\ pointed to S5~0716+714 twice, on 2004 April 4--5
and 2007 September 24--25 (Orbit 791 and 1427, PI: G. Tagliaferri),
lasting for $\sim59$~ks and $\sim74$~ks, respectively. The first
\xmm\ observation was analyzed by Ferrero et al. (2006, hereafter
FE06; see also Foschini et al. [2006]), who studied the temporal and
spectral variability of the synchrotron and IC component on
timescales of hours. S5~0716+714 was in a higher state compared to
previous observations. Strong variations with flux changes by a
factor of more than three were detected on timescale of hours. The
variability amplitude is significantly higher in the soft than in
the hard X-ray band. The soft X-ray variations correlate with the
harder X-ray ones, but no pronounced time lags were found. The
time-resolved spectral analysis with a double power law model showed
that the crossing points of the synchrotron and IC emission move to
higher energies with increasing fluxes. Both of the two components
vary on timescales of hours. The synchrotron emission becomes more
dominant when the source brightens, following a harder-when-brighter
trend as HBLs. The IC emission exhibits a more complicated
variability behaviour.

In this paper, we present in detail the temporal and spectral
variability of the synchrotron and IC emission for the second \xmm\
observation of S5~0716+714. Some new variability properties are
found. Our results will be compared with those of the first \xmm\
observation and the earlier observations by previous X-ray
telescopes. The X-ray observations are also compared with the
simultaneous optical and UV data obtained with \xmm.


\section{XMM-Newton Observation and Data Reduction}\label{sec:obs}

During \xmm\ (Jansen et al. 2001) flight of orbit 1427, S5~0716+714
was observed on September 24--25, 2007. EPIC-PN detector (Str${\rm
\ddot{u}}$der et al. 2001) was operated in imaging small window mode
with thin filter, while EPIC-MOS (MOS1 and MOS2) detectors (Turner
et al. 2001) did not collect data. Therefore, the X-ray data
analysis is performed only for the PN data. The Optical Monitor (OM;
Mason et al. 2001) was configured in standard imaging mode with four
different filters. We follow the standard procedures to reprocess
the Observation Data Files (ODF) with the \xmm\ Science Analysis
System (SAS) version 7.1.0 and with the latest calibration files as
of 2009 July 30.

We extract the high energy (10~keV~$<E<$~12~keV) PN light curve with
single event (PATTERN=0) only from the full frame of the exposed
CCD, to identify intervals affected by the high particle background.
We find that the PN detector encountered the high background from
$\sim45$~ks to the end of the observation. Pile-up effects are
examined with the SAS task {\it epatplot}, showing that the PN data
are not affected by such effects. As a result, we extract the PN
source counts from a circle region centered on the source with
radius of 36". The PN background counts are selected from a circle
region least affected by the source counts on the same CCD with the
same radius as the source region. Single and double events
(PATTERN=0--4) with quality FLAG=0 are selected for our analysis.
The count spectra are created with SAS task XMMSELECT, and grouped
with FTOOL task GRPPHA in order to have at least 30 counts in each
energy bin for the use of $\chi^2$ statistics. Redistribution
matrices and ancillary response files are produced with SAS task
RMFGEN and ARFGEN. The details of the PN exposure is summarized in
Table~\ref{tab:epic}.

Totally 54 exposures were obtained for OM observation, with an
exposure time of $800$~s for each exposure. In time order, the
filters were changed in the sequence V, U, UVW1 and UVM2, with 14,
15, 10 and 15 exposures, respectively. The standard imaging data
processing offers source count rates and instrumental magnitudes.
The results of OM observation are listed in Table~\ref{tab:om}.


\section{X-ray data analysis}\label{sec:xray}

\subsection{The average spectrum }\label{sec:spec}

In this section, we perform spectral fits to the average spectrum
extracted from the entire observation. The time intervals affected
by the high background are excluded when extracting the average
spectrum. The high background intervals are identified with the
10--12~keV light curve extracted from the full exposed CCD. The time
intervals with count rate larger than $0.1$ counts/s are identified
as the high background intervals.

Spectral fits are performed with XSPEC version 12.5.0. In order to
expediently compare to the results of the first \xmm\ observation,
we restrict the spectral fits in the same 0.5--10~keV band as used
in FE06. This is also in an attempt to exclude any possible
remaining (cross-)calibration uncertainties below 0.5~keV, since the
PN detector was operated in timing mode during the first \xmm\
observation. However, spectral fits in the 0.3--10~keV give similar
results. We adopt the same Galactic neutral hydrogen absorption
column density as used by FE06, i.e., $N_H=3.05\times10^{20}$~${\rm
cm^{-2}}$ directed to S5~0716+714 (Murphy et al. 1996). The Galactic
$N_H$ adopted by Foschini et al. (2006) is $3.81\times10^{20}$~${\rm
cm^{-2}}$ (Dickey \& Lockman 1990), slightly larger than the value
we used. The results of spectral fits with different models are
summarized in Table~\ref{tab:spec}.

\begin{figure}\epsscale{1}
\centering
\includegraphics[scale=0.41, angle=270]{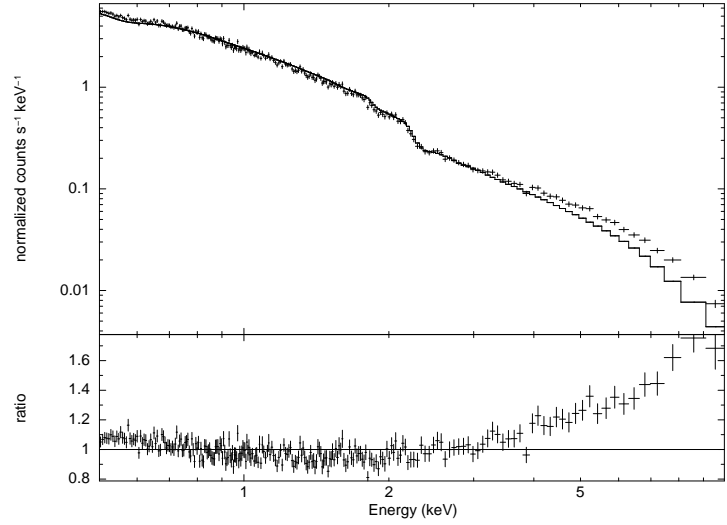}
\caption {\footnotesize The PN count spectrum and the best single
power law fit (top panel). The data-to-model ratios (bottom panel)
clearly show a spectral hardening above $\sim2$~keV.}
\label{fig:ratioPL}
\end{figure}

We first fit the spectrum with the model of a single power law plus
Galactic absorption to inspect whether the spectrum is typical of
LBLs, i.e., characterized by the concave shape. The count spectrum,
the best fit and the data-to-model ratios are shown in
Figure~\ref{fig:ratioPL}. The fit is obviously unacceptable (see
Table~\ref{tab:spec} for the fit statistics). The data-to-model
ratios are typical of LBLs, clearly showing the spectral hardening
above $\sim2-3$~keV. The source spectrum thus has a concave shape
typical of LBLs. A single power law with free Galactic absorption
does not present an acceptable fit either, giving rise to the $N_H$
value much lower than the Galactic one. This indicates that the
concave spectrum is intrinsic to the source. Therefore, we fix $N_H$
to the Galactic value in all of the following spectral fits.

The broken power law is frequently used to describe the X-ray
spectra of BL Lac objects, irrespective of a convex (for HBLs) or a
concave (for LBLs) shape. We therefore use this model to fit the
average spectrum of S5~0716+714 and obtain a good fit. The best fit
yields the concave spectrum, with photon indices of
$\Gamma_1=2.38\pm0.03$ and $\Gamma_2=1.92^{+0.03}_{-0.07}$ below and
above the break energy of $E_{\rm break}=1.94^{+0.32}_{-0.11}$~keV.
The spectral hardening of $\Delta\Gamma \sim0.46$, along with the
break energy and photon indices are typical of LBLs. The
data-to-model ratios are shown in the first panel of
Figure~\ref{fig:ratio}. Although the fit is good to be acceptable,
it is worth noting that the fit leaves a visible hard tail above
$\sim8$~keV, indicating a further spectral hardening.

As the X-ray spectra of LBLs are thought to be composed of a steep
synchrotron and a flat IC component, it is reasonable to use the
double power law to fit the spectrum of S5~0716+714. In this model,
the steeper power law describes the synchrotron component, while the
flatter power law represents the IC component (e.g., FE06). The fit
is excellent. The data-to-model ratios are shown in the second panel
of Figure~\ref{fig:ratio}, from which one can see that the hard tail
left from the broken power law fit, disappears. This suggests that
the double power law presents a better fit than the broken power law
does. The fit statistics ($\chi^2_{\nu}$ and probability,
Table~\ref{tab:spec}) do show the fit improvement. The best fit
gives the steeper and the flatter photon indices as
$\Gamma_1=2.56^{+0.09}_{-0.07}$ and $\Gamma_2=1.22^{+0.19}_{-0.21}$,
respectively.

The logarithmic parabola with the form of $E^{-(\Gamma+\beta
logE)}$, often used to describe continuously downward-curved
(convex) X-ray spectra of HBLs (e.g., Massaro et al. 2004a; Zhang
2008), also presents a good fit to the spectrum. The curvature index
$\beta$ is $-0.37\pm0.02$, suggesting that the spectrum is upward
curved, opposite to HBLs. The photon index at 1~keV is $\Gamma =
2.37\pm0.01$, similar to the value of the soft photon index in the
broken power law. The implication of the logarithmic parabolic model
is consistent with that of the broken power law model, but it
presents a more precise description for the continuously flattening
of the spectrum. The data-to-model ratios are shown in the third
panel of Figure~\ref{fig:ratio}. The residual hard tail is weaker
than that left from the broken power law. Although it presents a
better fit than the broken power law, the logarithmic parabola is
not good as the double power law. Moreover, the physical
implications of the logarithmic parabola are not straightforward to
LBLs, so we do not discuss it any more.

Finally, a power law plus a black body or a thermal bremsstrahlung
component also provides acceptable fits. The data-to-model ratios
are shown in the fourth and fifth panel of Figure~\ref{fig:ratio}.
Like the broken power law, the two models leave similar hard tails.
As indicated by the statistic parameters in Table~\ref{tab:spec},
the fits with the two models is better than the broken power law
fit, but they are worse than the fits with the double power law and
the logarithmic parabola. However, as discussed in FE06, the two
models can not be easily related with any physical interpretation
for the source. Therefore, despite both of the models provide
adequate representations for the spectrum, they are not discussed
any more either. We present the fit results of the two models for
the purpose of comparisons with those obtained by FE06 for the first
\xmm\ observation of the source. Basically, the fit parameters are
similar for the two observations.

In section~\ref{sec:timespec}, we adopt only the broken power law
and the double power law as our favorite models to study spectral
variability of S5~0716+714. The broken power law provides a simple
and straightforward parameter representation for the concave X-ray
spectrum of S5~0716+714, in which the soft and hard X-ray emissions
are physically dominated by the high energy tail of the synchrotron
emission and the low energy side of the IC emission, respectively.
Nevertheless, the double power law disentangles directly the concave
X-ray spectrum into the steeper and flatter power law components,
implicitly associating them with the high energy tail of the
synchrotron emission and the low energy side of the IC emission,
respectively. Furthermore, the contributions of the two emission
components to the total fluxes can be qualitatively estimated for
any energy bands. For example, the synchrotron and IC emission
contribute to $\sim65.79\%$ and $\sim34.21\%$ of the 0.5--10~keV
flux, respectively, similar to those obtained for the first \xmm\
observation (FE06). In the 0.5--2~keV band, the synchrotron
component contributes to most (up to 85.95\%) of the flux. In the
2--10~keV band, however, it contributes to only 44.85\% of the flux,
smaller than the contribution of the IC component.


\subsection{Temporal Variability}\label{sec:lc}

\begin{figure}\epsscale{0.5}
\plotone{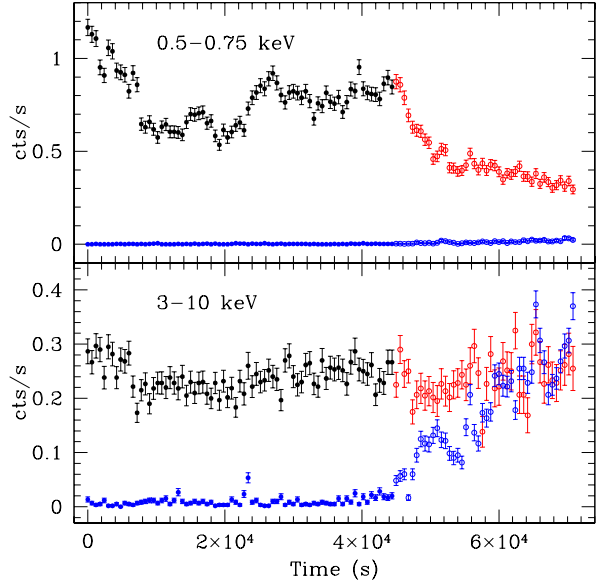} \caption { \footnotesize The top panel shows
the background subtracted source (black and red) and background
(blue) light curve in the 0.5--0.75~keV band, respectively. The
bottom panel plots the background subtracted source light curve
(black and red) and background light curve (blue) in the 3--10~keV
band, respectively. The bin size is 600~s and the time zero is
JD=2454368.21282. The background light curves clearly show that the
high background, starting from $\sim45$~ks to the end of the
observation, is mostly significant in the high energy band. The open
symbols indicate the high background interval. }
\label{fig:lc:softhard}
\end{figure}

\begin{figure}\epsscale{0.45}
\plotone{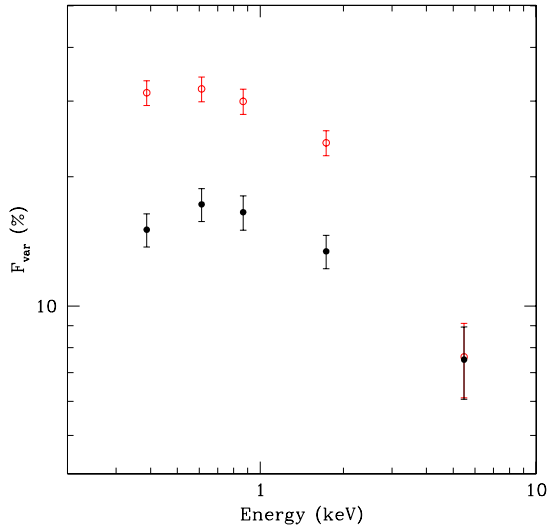} \caption { \footnotesize The variability amplitude
(\fvar) as a function of photon energy. The black solid symbols are
calculated by excluding the high background interval, while the red
open symbols are calculated with the entire observation. }
\label{fig:fvar}
\end{figure}

\begin{figure}\epsscale{0.45}
\plotone{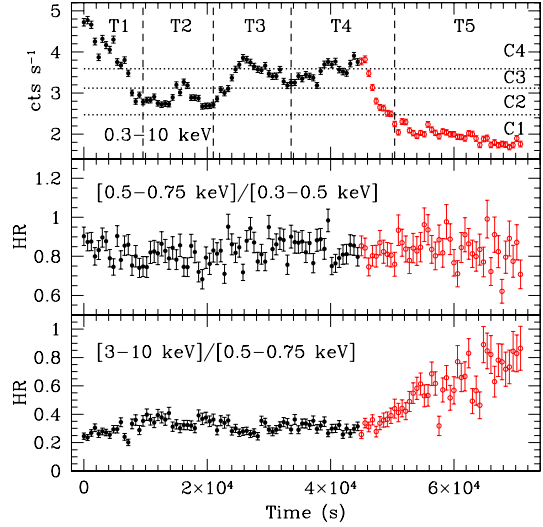} \caption { \footnotesize From top to bottom panels
are shown the background subtracted 0.3--10~keV light curve, and the
temporal evolution of the 0.5--0.75 to 0.3--0.5~keV hardness ratios
and of the 3--10 to 0.5--0.75~keV hardness ratios, respectively. The
top panel also shows the time intervals (the vertically dashed
lines) used for the time resolved spectral analysis, and the count
rate intervals (the horizontally dotted lines) used for the count
rate resolved spectral analysis (see Section~\ref{sec:timespec}).
The data have a binsize of 600~s. The red open circles show the time
interval affected by the high background.} \label{fig:lchr}
\end{figure}

\begin{figure}\epsscale{1}
\plottwo{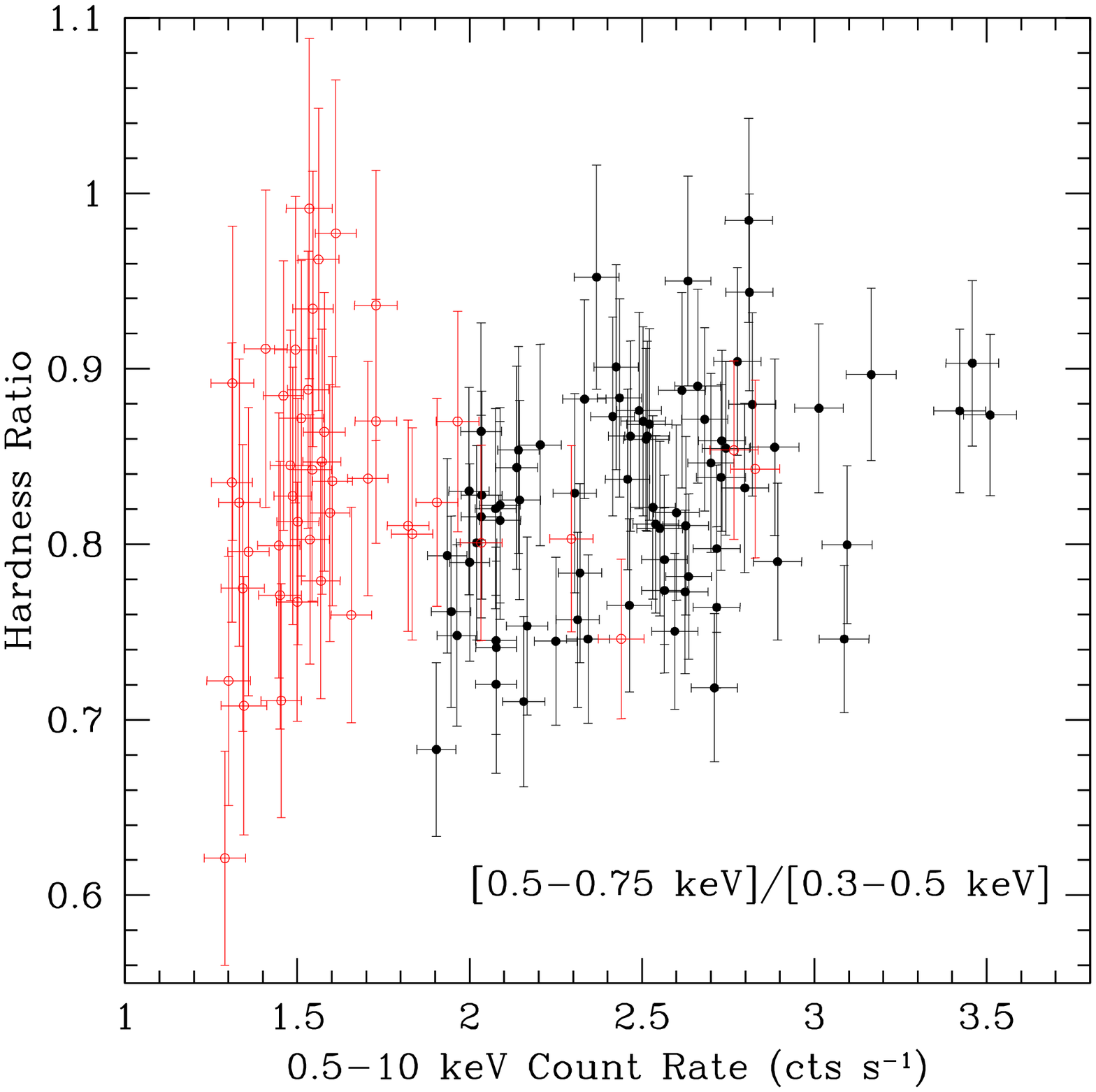}{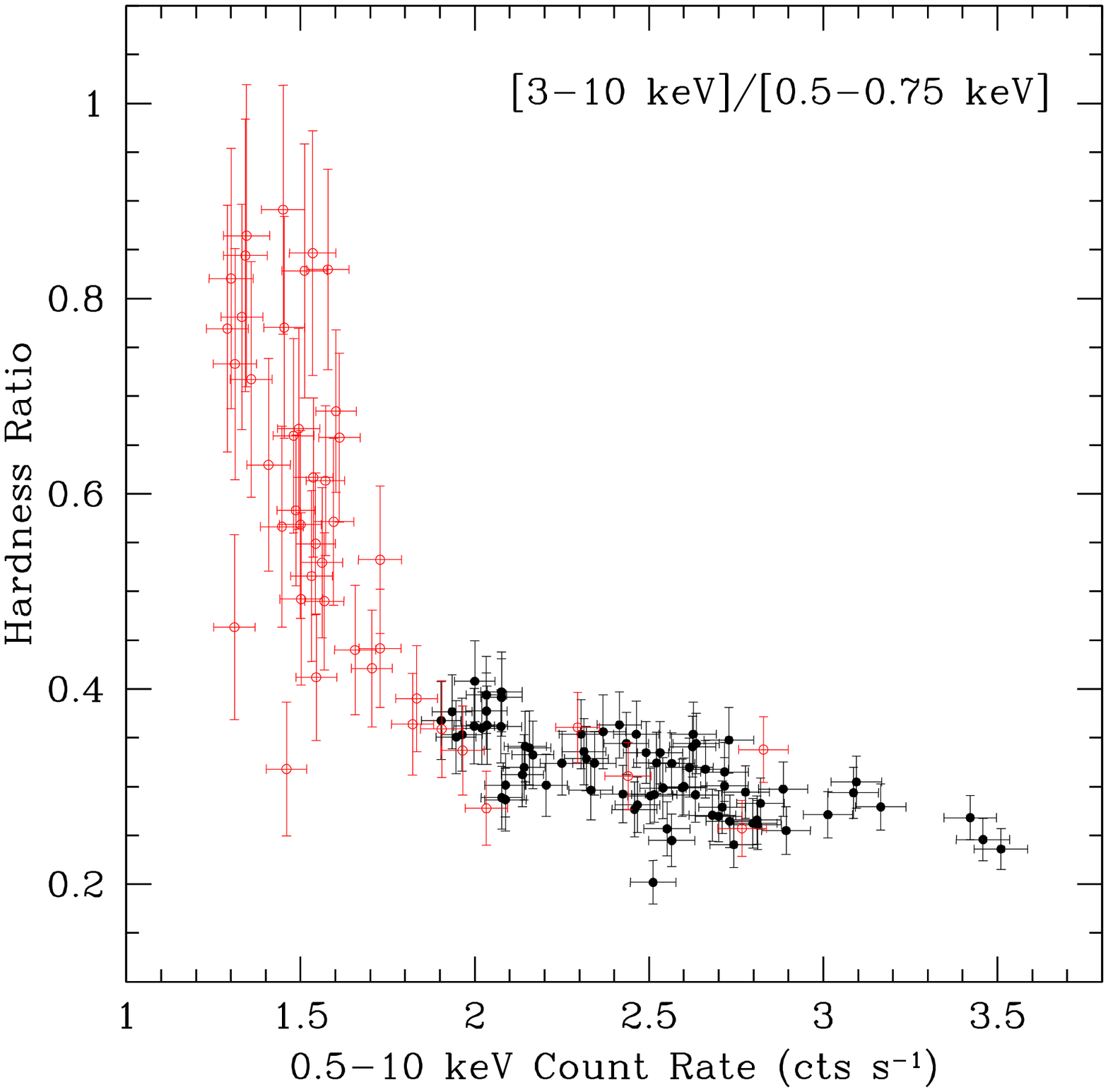} \caption { \footnotesize Relationship
between the hardness ratios and the count rates. Left plot is for
the 0.5--0.75 to 0.3--0.5~keV hardness ratios versus the 0.5--10~keV
count rates. Right plot is for the 3--10 to 0.5--0.75~keV hardness
ratios versus the 0.5--10~keV count rates. The data have a time
binszie of 600~s. The red open circles indicate the data polluted by
the high background. } \label{fig:loop}
\end{figure}

Figure~\ref{fig:lc:softhard} shows the 600~s binned background
subtracted light curve in the 0.5--0.75 and 3--10~keV band,
respectively. The corresponding background light curves are also
plotted for comparisons, clearly showing that the background is
significantly high from the $\sim45$~ks to the end of the
observation. The background, however, is high in the hard X-ray band
only. In order to study the entire observation, we discriminate the
time interval affected by the high background with open symbols in
Figures~\ref{fig:lc:softhard}--\ref{fig:loop}.

The 0.5--0.75~keV light curve illustrates that the source is
strongly variable in the soft X-ray band. The variability of a
factor of $\sim4$ is quantified by the ratio of the maximum to
minimum count rates over the whole observation. The light curve is
plenty of characteristics. It initially decreases by a factor of
$\sim2$ over a timescale of $\sim8$~ks (i.e., the halving timescale
is $\sim8$~ks), followed by a small amplitude flare with duration of
$\sim6$~ks. Afterwards, a large amplitude flare occurs. The count
rates increase by a factor of $\sim1.5$ on a timescale of
$\sim7$~ks. Then another flare of similar amplitude appears to
happen, whose rising phase may overlap with the decaying phase of
its anterior flare. Starting from the $\sim46$~ks of the
observation, the source decreases rapidly by a factor of $\sim2$
over a timescale of $\sim7$~ks, followed by a slow decay till the
end of the observation.

However, the 3--10~keV light curve shows only a weak trend of
variability, with roughly $50\%$ changes from the minimum to maximum
count rate. From the beginning to $\sim50$~ks of the observation,
the hard band light curve approximately follows the soft band light
curve. Nevertheless, during the last part of the observation, the
hard band light curve becomes erratic, probably caused by the high
background, and does not track the soft band light curve.

Accordingly, S5~0716+714 is much more variable in the soft than in
the hard X-ray band. We use the factional variability amplitude
(\fvar, e.g., Zhang et al. 1999) to quantify the source's
variability. In the 0.5--10~keV band, \fvar\ is ($25.2\pm1.6$)\%. In
order to study the energy dependence of the variability amplitude,
we calculate \fvar\ in five (0.3--0.5, 0.5--0.75, 0.75--1, 1--3 and
3--10~keV) energy bands. The results are tabulated in
Table~\ref{tab:fvarlag}. In Figure~\ref{fig:fvar}, \fvar\ is plotted
against the photon energy, where the solid and open symbols indicate
the results by excluding and including the high background interval,
respectively. Except for the different values of \fvar, the energy
dependence of \fvar\ is similar for the two scenarios. \fvar\
increases from the 0.3--0.5~keV to the 0.5--0.75~keV, then it
decreases with higher energies.

In Figure~\ref{fig:lchr}, we plot, from the top to bottom panels,
the 0.3--10~keV light curve, and the temporal evolution of the
0.5--0.75 to 0.3--0.5~keV hardness ratios (representing the soft
X-ray spectra) and the 3--10 to 0.5--0.75~keV hardness ratios
(representing the overall X-ray spectra), respectively. The
0.5--0.75 to 0.3--0.5~keV ratios appear to follow the light curve,
in the sense that the soft X-ray spectra flatten with increasing
fluxes. On the contrary, the 3--10 to 0.5--0.75~keV ratios seem to
anti-correlate with the light curve, indicating that the overall
spectra soften with increasing fluxes.

The left plot of Figure~\ref{fig:loop} shows the relationship
between the 0.5--0.75 to 0.3--0.5~keV ratios and the 0.5--10~keV
count rates. Except for the data contaminated by the high
background, the hardness ratios appear to increase with higher count
rates, showing the harder-when-brighter trend for the soft X-ray
variations. The right plot of Figure~\ref{fig:loop} presents the
relationship between the 3--10 to 0.5--0.75~keV ratios and the
0.5--10~keV count rates. It is clear that the hardness ratios become
smaller with higher count rates, indicating the softer-when-brighter
phenomenon for the overall X-ray variations.

\begin{figure} \epsscale{0.6}
\plotone{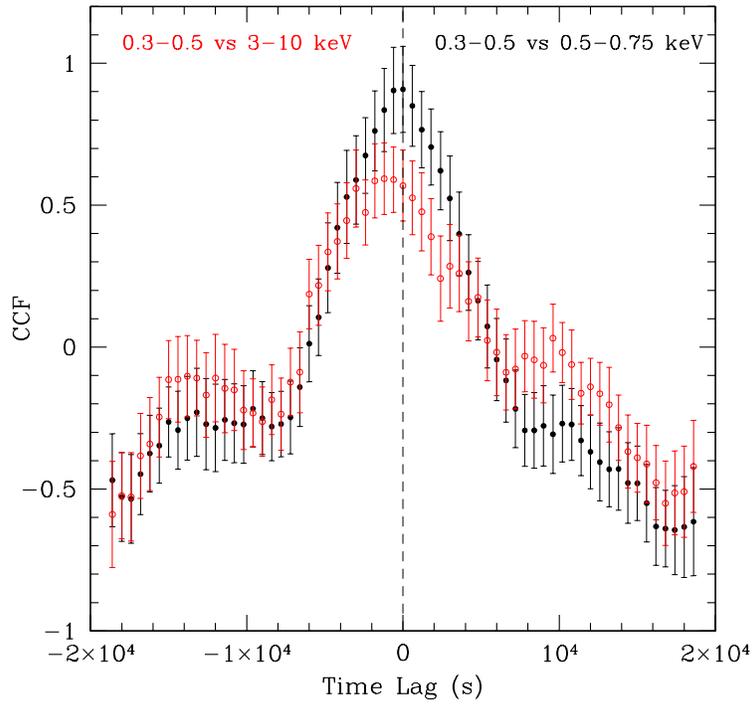} \caption {\footnotesize The black solid circles
show the CCF between the 0.3--0.5 and 0.5--0.75~keV light curve,
peaking at zero lag. The red open circles present the CCF between
the 0.3--0.5 and 3--10~keV light curve, peaking at $-1200$~s. The
negative lags indicate a delay of the lower energy with respect to
the higher energy light curve. The light curves and the CCFs are
binned over 600~s. The high background interval is excluded when
calculating the CCFs. } \label{fig:ccf}
\end{figure}

Due to the long orbital period, \xmm\ is able to produce
continuously sampled light curves over about one day, which is very
important to study the inter-band time lags of the intra-day X-ray
variability of blazars. The lags can thus be estimated by
calculating the standard Cross-Correlation Function (CCF) between
any two energy band light curves. We calculate the CCFs between the
0.3--0.5~keV and the higher energy light curves. The time interval
affected by the high background are excluded. Figure~\ref{fig:ccf}
plots the CCFs of the 0.3--0.5~keV with respect to the 0.5--0.75 and
3--10~keV band, respectively. A negative lag indicates that the
lower energy variations lag the higher energy ones (i.e., soft lag).
The CCF of the 0.3--0.5~keV versus the 0.5--0.75~keV peaks at zero
lag, but the slight asymmetry towards negative lags suggests a
possible soft lag of $<600$~s. However, the CCF between the 0.3--0.5
and 3--10~keV peaks at the lag of $-1200$~s, suggesting a soft lag
of $\sim 1200$~s. We estimate the lags with three techniques: (1)
$\tau_{\rm peak}$ is the lag corresponding to the CCF peak ($\rm
CCF_{\rm max}$); (2) $\tau_{\rm cent}$ is obtained by computing the
CCF centroid; and (3) $\tau_{\rm fit}$ is the lag corresponding to
the peak of a Gaussian function (plus a constant) fitted to the CCF.
Both $\tau_{\rm cent}$ and $\tau_{\rm fit}$ are measured within the
CCF lag range of $[-7000, 7000]$~s only. The results are tabulated
in Table~\ref{tab:fvarlag}, where the uncertainties on $\tau_{\rm
fit}$ are $1 \sigma$ confidence level. Although $\tau_{\rm peak}$ is
zero for the CCFs of the 0.3--0.5~keV with respective to the
0.5--0.75, 0.75--1 and 1--3~keV, the estimated values of $\tau_{\rm
cent}$ and $\tau_{\rm fit}$ suggest soft lags of $\sim200-600$~s,
smaller than the binsize (600~s) of the light curves. This means
that for the three cases, the CCFs may peak at lags of 0--600~s
rather than at zero if the light curves were binned with smaller
binsizes (e.g., 100~s). Interestingly, from the point view of
$\tau_{\rm cent}$ or $\tau_{\rm fit}$, the soft lags become larger
with larger energy differences.


\subsection{Spectral Variability}\label{sec:timespec}

In order to investigate the spectral variability of the synchrotron
and IC emission and their sum, respectively, we divide the entire
observation into several intervals in two ways. The first division
is by time order. We identify the 0.3--10~keV light curve as five
"isolated" episodes, separated with the vertically dashed lines in
the top panel of Figure~\ref{fig:lchr} and numbered as T1, T2, T3,
T4 and T5 in time order. T1 may represent a decay phase (possibly
part) of a large amplitude flare. T2 covers mainly a low amplitude
flare. T3 and T4 present two pronounced flares. The decaying phase
of T3 and the rising phase of T4 may overlap with each other. T4 is
complicated, consisting of a couple of small amplitude flares. T5 is
a long slow decay. Then we divide the observation on the basis of
the 0.5--10~keV count rates. We choose four count rate intervals,
differentiated with the horizontally dotted lines in the top panel
of Figure~\ref{fig:lchr} and numbered as C1, C2, C3 and C4 from the
low to high count rates. C4 is the highest state, covering the peaks
of T1, T3 and T4. We mention that C1 is roughly identical to T5, as
the quiescent state in both divisions. Since they are affected by
the high background, we separate the T5 and C1 intervals from other
intervals. The T5 and C1 spectra might be distorted by the high
background.

We repeat spectral fits for each time and count rate interval with
the broken and double power law plus the Galactic absorption. The
broken power law presents a visual inspection of the break energy
from the synchrotron to IC dominance. The double power law
determines the crossing energy, at which the fluxes of the two
emission components are equal. Our main purpose is to study the
changes of the break and crossing energies with the source's
intensities. The results of the spectral fits are summarized in
Table~\ref{tab:fit:bknpl} and Table~\ref{tab:fit:2pl} for the broken
and double power law, respectively, showing that all fits are
acceptable. In most cases, however, the fits are somewhat better for
the double power law than for the broken power law. It is important
to note that the crossing energies, depending on the photon indices
and the normalization of the two components, are not identical to
the break energies in general.

\begin{figure}\epsscale{1}
\plottwo{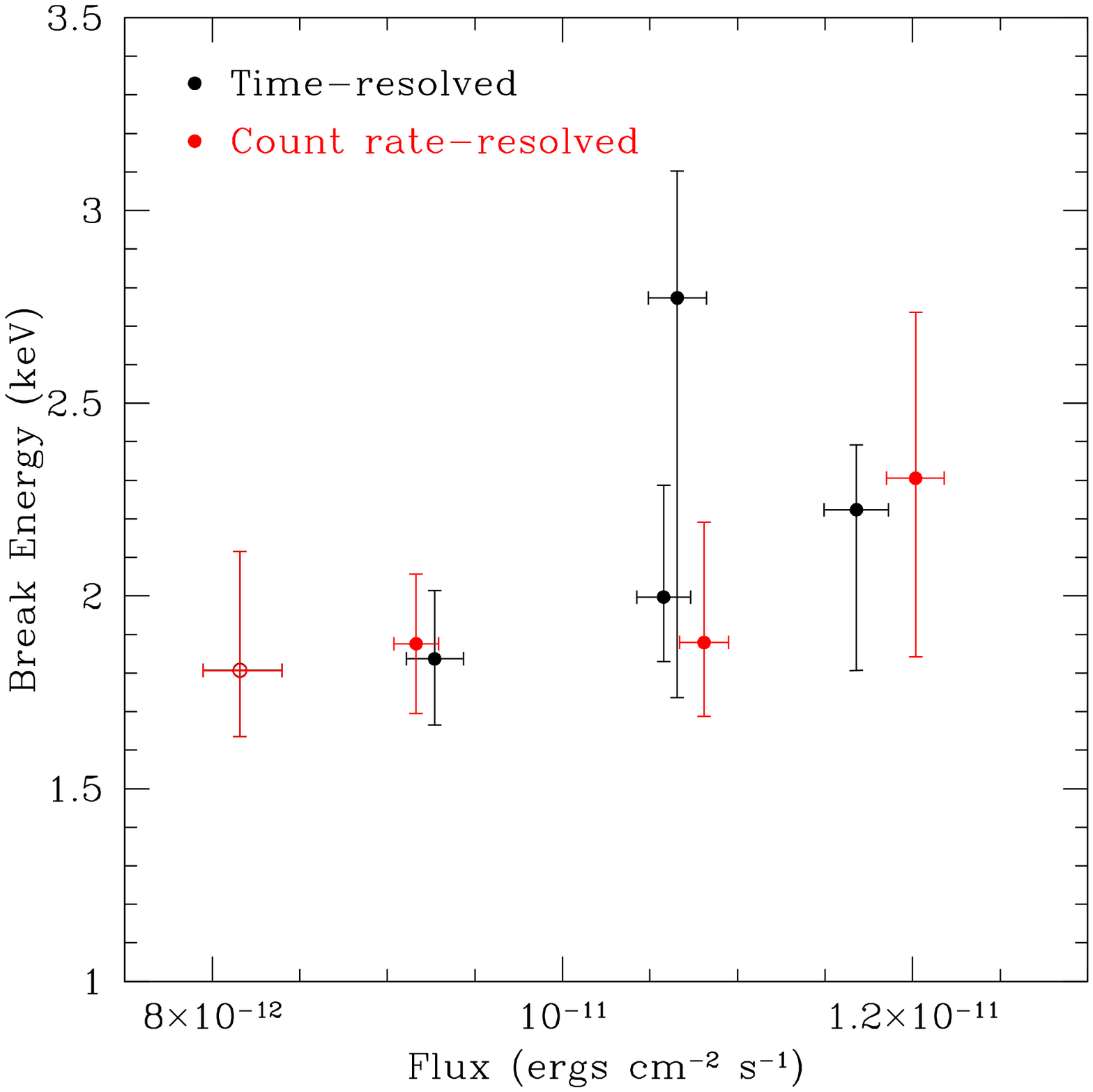}{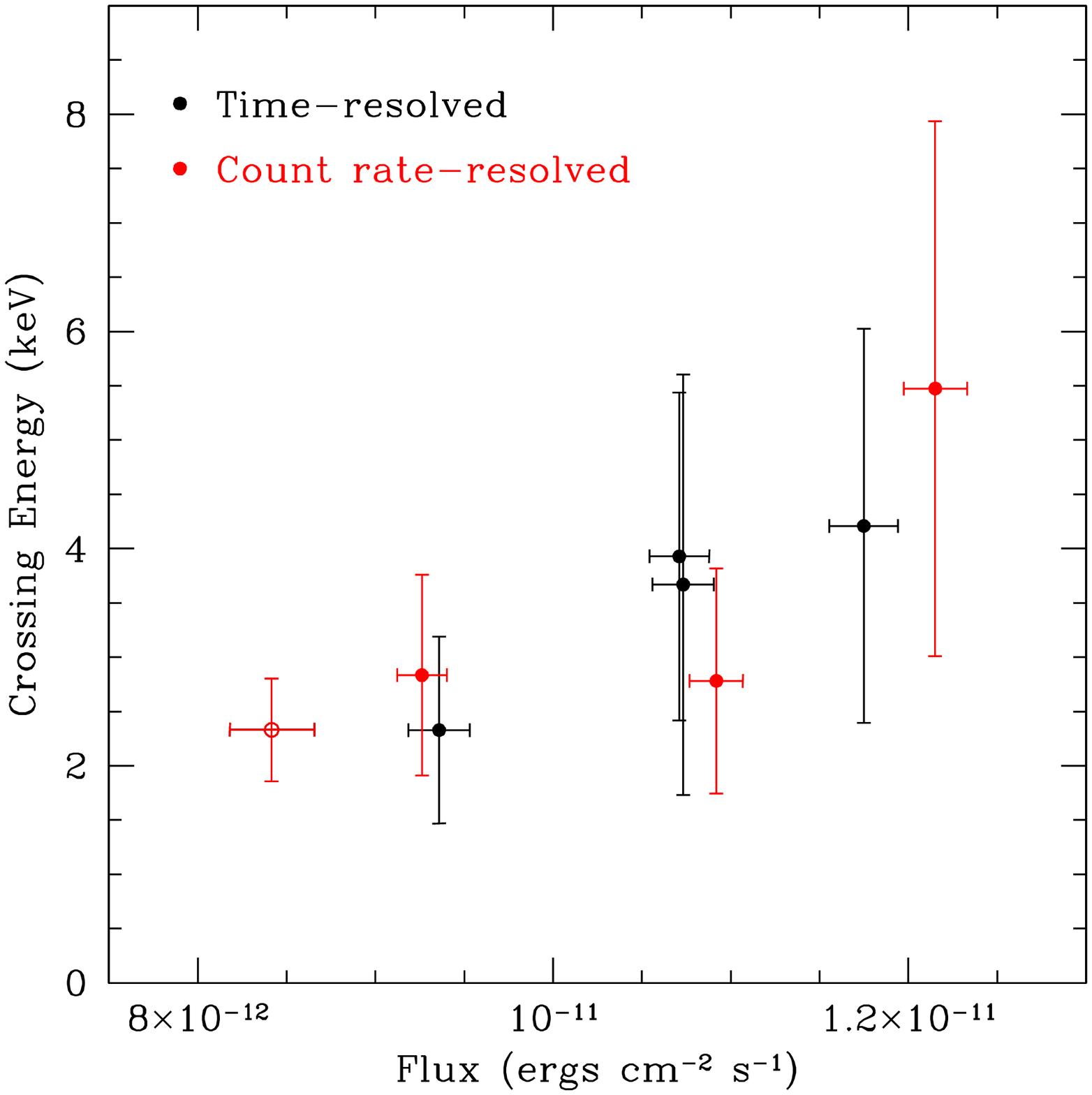} \caption { \footnotesize The left
plot shows the relationship between the break energies and the
0.5--10~keV total fluxes, derived from the broken power law fit. The
right plot shows the relationship between the crossing energies (of
the synchrotron and IC components) and the 0.5--10~keV total fluxes,
derived from the double power law fit. The black symbols are for the
time-resolved intervals, and the red symbols for the count
rate-resolved intervals. The open symbol indicates the T5/C1
interval together, which is affected by the high background. }
\label{fig:ecross}
\end{figure}

Figure~\ref{fig:ecross} plots the break and crossing energies as a
function of the 0.5--10~keV total fluxes. The two kinds of energies
increase when the fluxes increase, but the crossing energies appear
to increase more rapidly with the fluxes than the break energies
does.

\begin{figure}\epsscale{1}
\plottwo{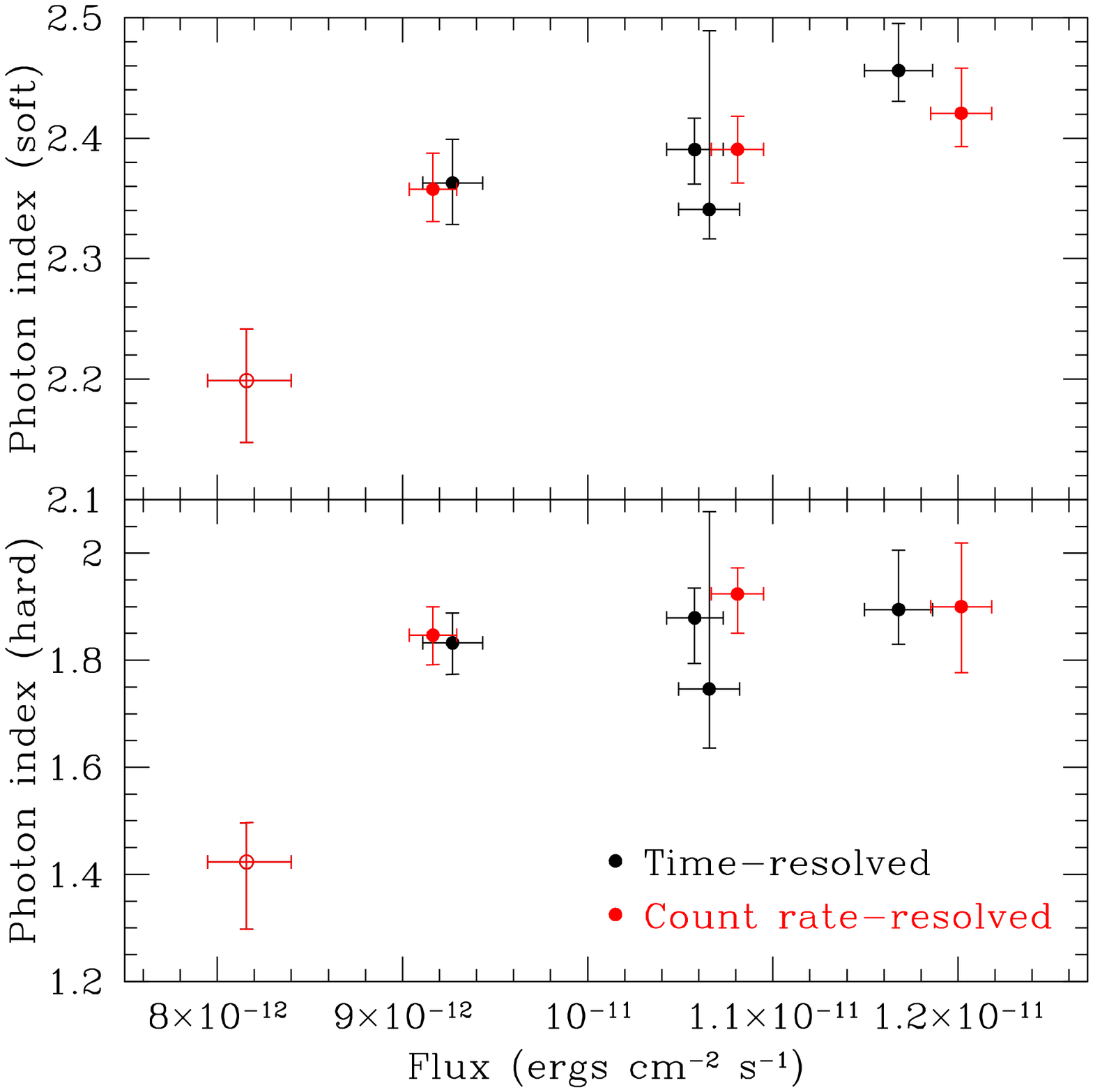}{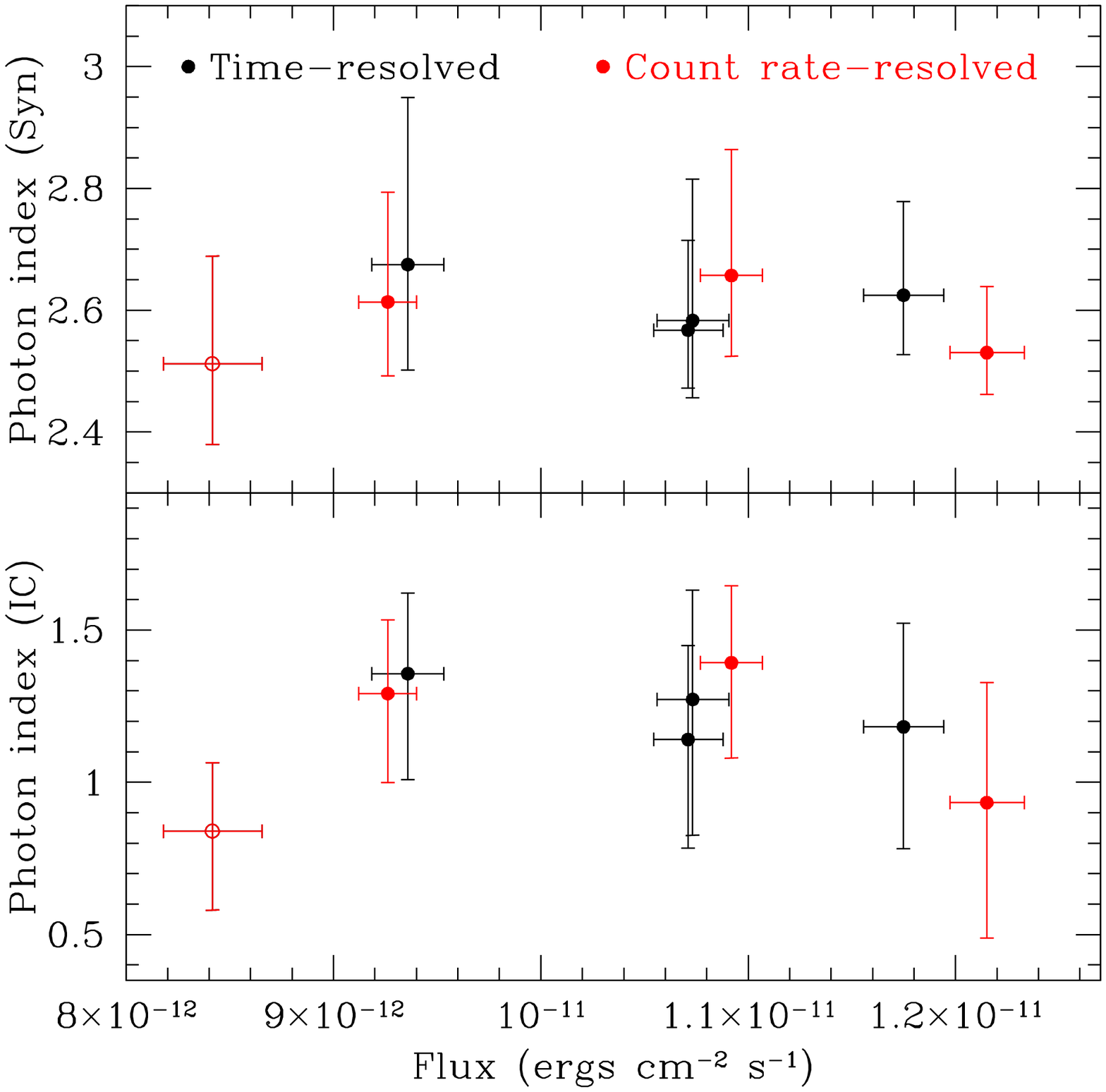} \caption {\footnotesize The
photon indices are plotted against the 0.5--10~keV total fluxes. The
left plot shows the soft and hard X-ray photon indices, derived from
the broken power law fit. The right plot shows the synchrotron (Syn)
and IC photon indices, derived from the double power law fit. The
black symbols are for the time-resolved intervals, and the red
symbols for the count rate-resolved intervals. The open symbol
indicates the T5/C1 interval together, which is affected by the high
background.} \label{fig:index}
\end{figure}

In Figure~\ref{fig:index}, we show the relationships between the
photon indices and the 0.5--10~keV total fluxes. For the broken
power model, there is weak correlation, showing that the soft photon
indices become larger (the spectra become steeper) with increasing
fluxes. However, except for the T5/C1 point, the hard photon indices
are roughly identical within the uncertainties. It is worth noting
that the break energies change by $\sim 0.5$~keV between different
intervals, indicating that the soft and hard photon indices are
actually measured in different energy ranges. Therefore, the
implications for the changes of the soft and hard photon indices are
somewhat vague. For the double power law model, the synchrotron and
IC photon indices do not correlate with the fluxes. In fact, they
are consistent with each other within the error bars, no matter the
synchrotron or IC photon indices.

\begin{figure}\epsscale{0.45}
\plotone{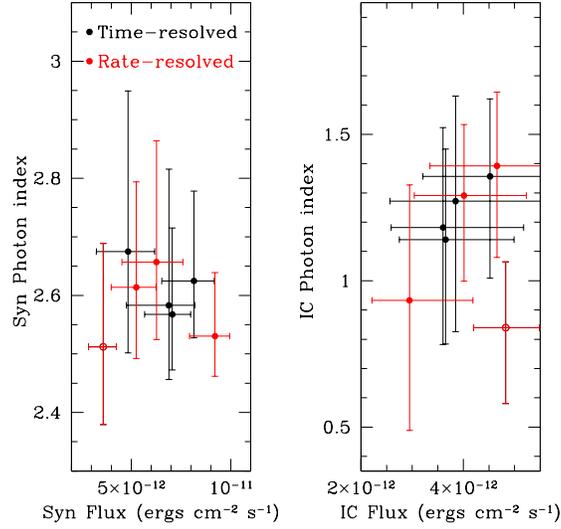} \caption {\footnotesize The left panel plots
the relationship between the synchrotron (Syn) photon indices and
the 0.5--10~keV synchrotron fluxes. The right panel plots the
relationship between the IC photon indices and the 0.5--10~keV IC
fluxes. The black symbols are for the time-resolved intervals, and
the red symbols for the count rate-resolved intervals. The open
symbol indicates the T5/C1 interval together, which is affected by
the high background. Besides the T5/C1 whose spectrum might be
affected by the high background, the synchrotron spectra appear to
harden with higher synchrotron fluxes, while the IC spectra appear
to soften with higher IC fluxes. } \label{fig:synicindex}
\end{figure}

The left panel of Figure~\ref{fig:synicindex} plots the relationship
between the synchrotron photon indices and the 0.5--10~keV
synchrotron fluxes. Although they might be identical within the
uncertainties, the synchrotron photon indices appear to become
smaller (i.e., the synchrotron spectra harden) with higher
synchrotron fluxes. The right panel of Figure~\ref{fig:synicindex}
shows the relationship between the IC photon indices and the
0.5--10~keV IC fluxes. Although their uncertainties are large, the
IC photon indices exhibit larger changes with respect to the
synchrotron ones. Contrary to the synchrotron spectral behaviour,
the IC photon indices appear to become larger (i.e., the IC spectra
soften) with larger IC fluxes.

\begin{figure}\epsscale{0.45}
\plotone{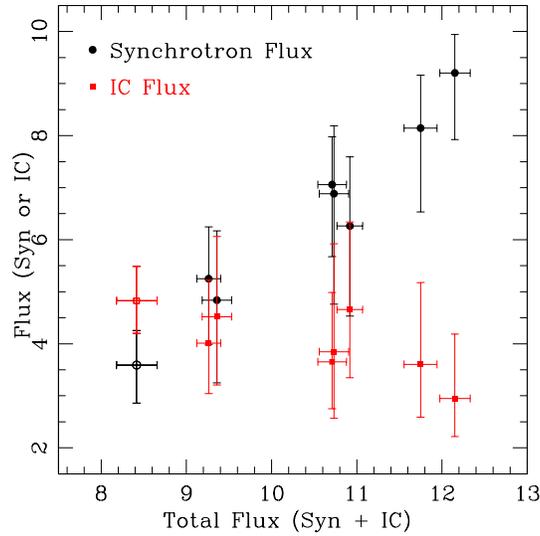} \caption {\footnotesize The synchrotron
(Syn) and IC 0.5--10~keV fluxes are plotted against the total
(synchrotron plus IC) 0.5--10~keV fluxes. The fluxes are in unit of
$\rm{10^{-12}~ergs ~ cm^{-2}~ s^{-1}}$. The open symbol indicates
the T5/C1 interval together, which is affected by the high
background. } \label{fig:synicflux}
\end{figure}

In Figure~\ref{fig:synicflux}, the synchrotron and IC 0.5--10~keV
fluxes are plotted against the total (synchrotron plus IC)
0.5--10~keV fluxes, respectively. With increasing total fluxes, the
synchrotron fluxes increase but the IC fluxes decrease, indicating
that the synchrotron fluxes anti-correlate with the IC fluxes.
Figure~\ref{fig:synicflux} also shows that the synchrotron fluxes
have larger changes than the IC fluxes. Similar relationships can be
obtained if the synchrotron and IC 0.5--2~keV (or 2--10~keV) fluxes
are plotted against the total 0.5--10~keV (or 0.5--2~keV, 2--10~keV)
fluxes, respectively.

\begin{figure}\epsscale{1}
\plottwo{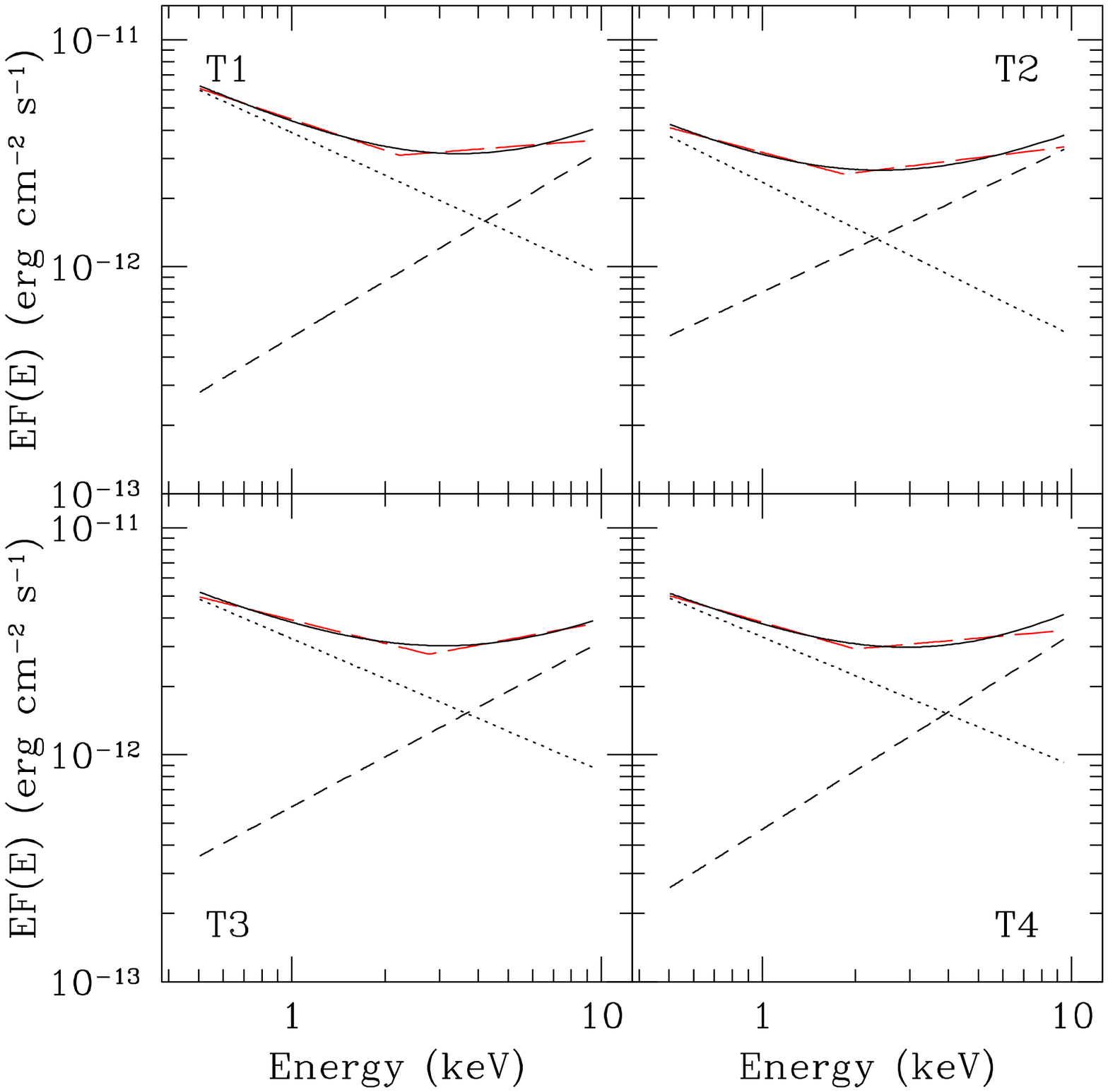}{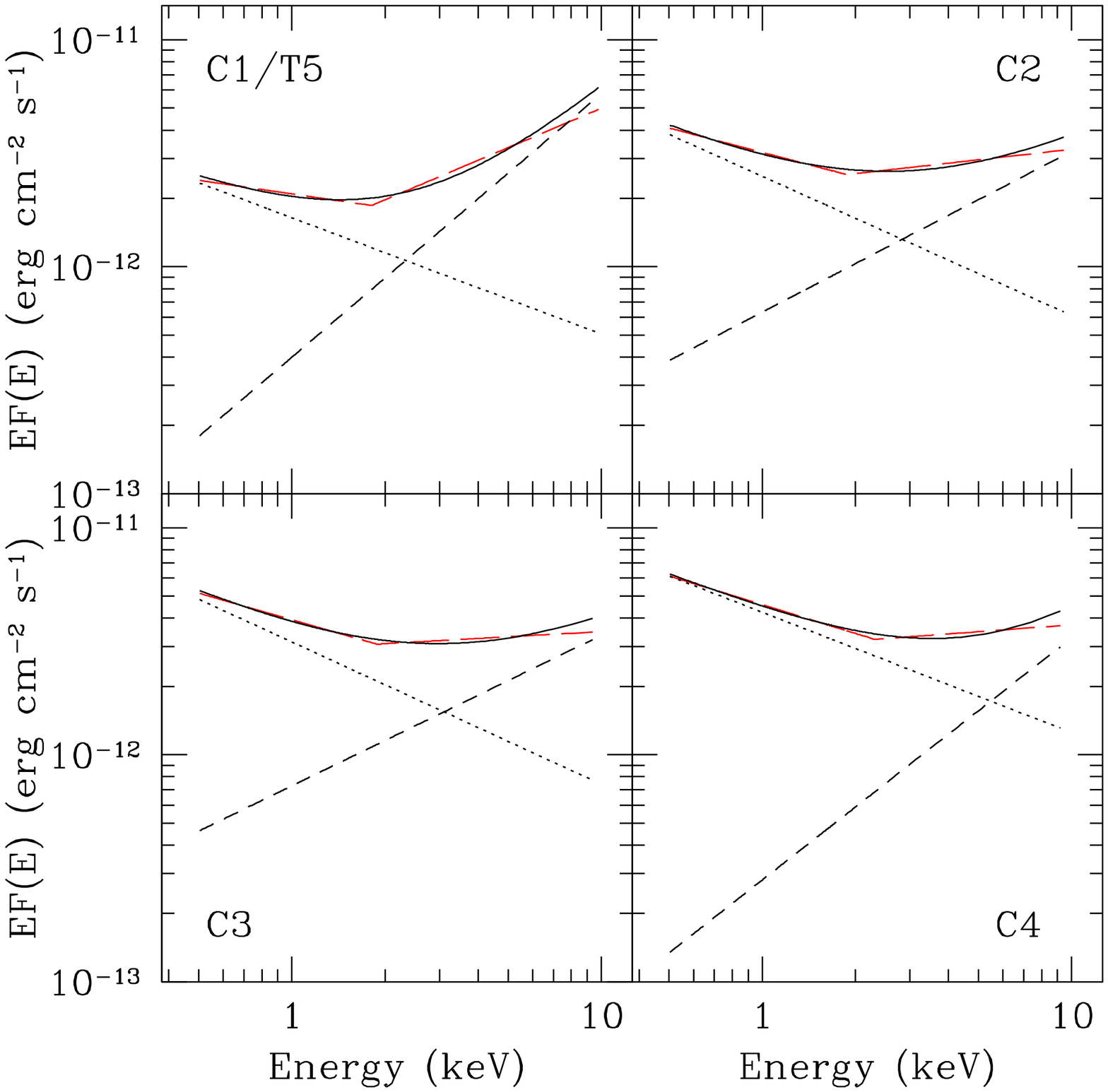} \caption
{\footnotesize The evolution of the unabsorbed X-ray SEDs. The black
dotted, short dashed and solid lines show the synchrotron and IC
component and the sum of the two, respectively, unfolded with the
double power law. The red long dashed line shows the SEDs unfolded
with the broken power law. The T5 and C1 interval are identical,
shown as one, which are affected by the high background.}
\label{fig:ecrossxsed}
\end{figure}

The evolutions of the unabsorbed 0.5--10~keV X-ray SEDs are shown in
Figure~\ref{fig:ecrossxsed}. We plot the SEDs unfolded with the
double and broken power laws together to show their differences. The
SED differences between the two models are visible, as viewed by the
disagreements between the SED break energies (the broken power law)
and the SED trough energies (the double power law). Their SEDs also
show clear differences in the highest energy band.

In Figure~\ref{fig:ecrossxsed}, the synchrotron and IC SEDs are also
plotted to show the evolutions of the two components separately and
their crossing energies. The crossing energies are obviously larger
than the break energies. The SED (double power law) trough energies
are smaller than the crossing energies in most cases. When the
source brightens, the synchrotron emission extends to higher
energies, whereas the IC emission recedes from lower energies.
Moreover, with higher total fluxes, the synchrotron fluxes have
larger increases than the IC fluxes. At the same same, the
synchrotron spectra harden with higher synchrotron fluxes, while the
IC spectra soften with larger IC fluxes (see also
Figure~\ref{fig:synicindex}). Therefore, the crossing energies
become larger when the source brightens (see also the right plot of
Figure~\ref{fig:ecross}). It is important to note that the
extensions of the synchrotron component to higher energies are
mainly caused by the significant increases of its normalization
rather than by its spectral hardening. The changes of the
synchrotron spectral slopes are in fact small. The normalization
changes of the IC component are small, while the changes of its
slopes are relatively large.

\begin{figure}\epsscale{1}
\plottwo{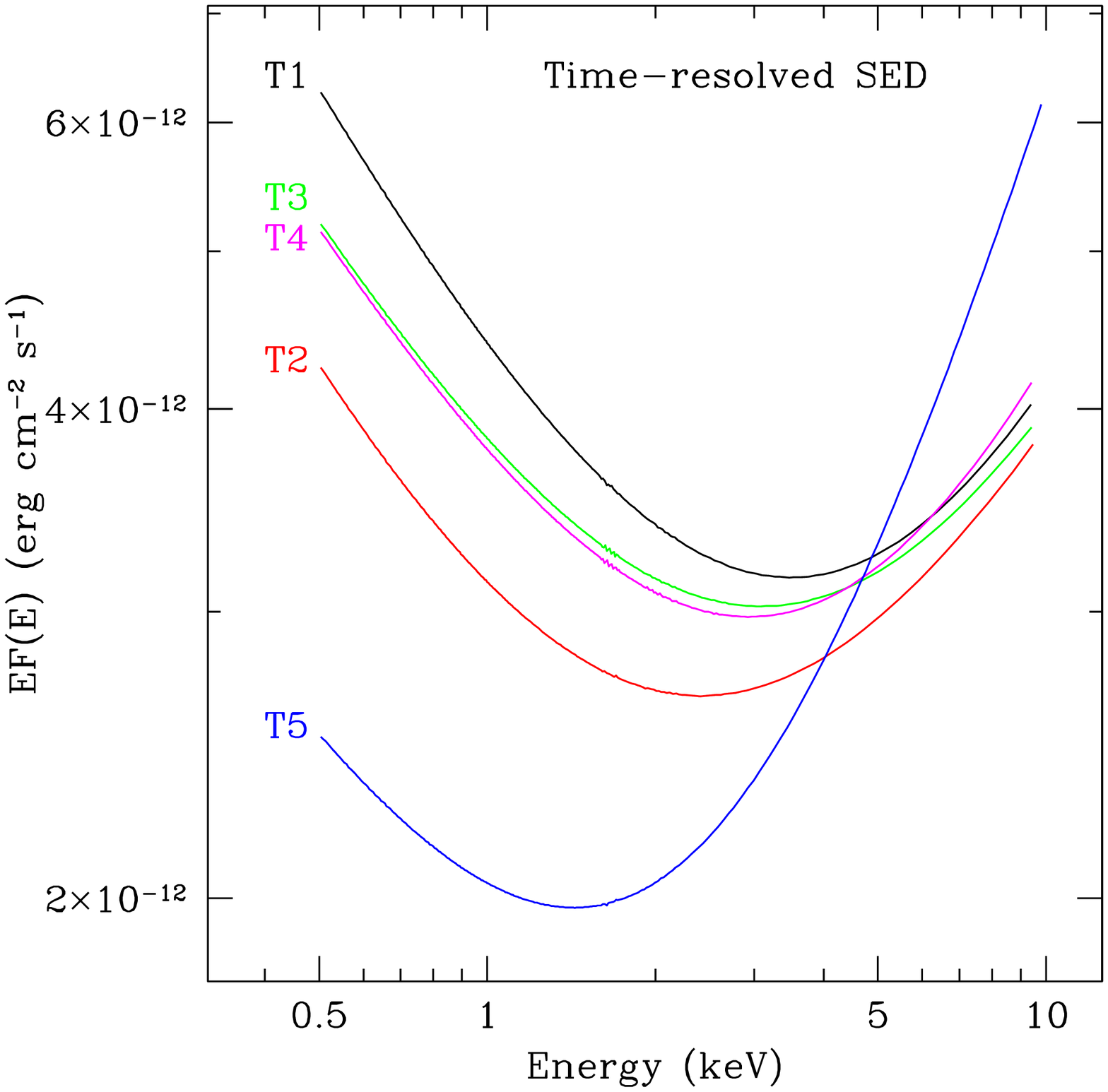}{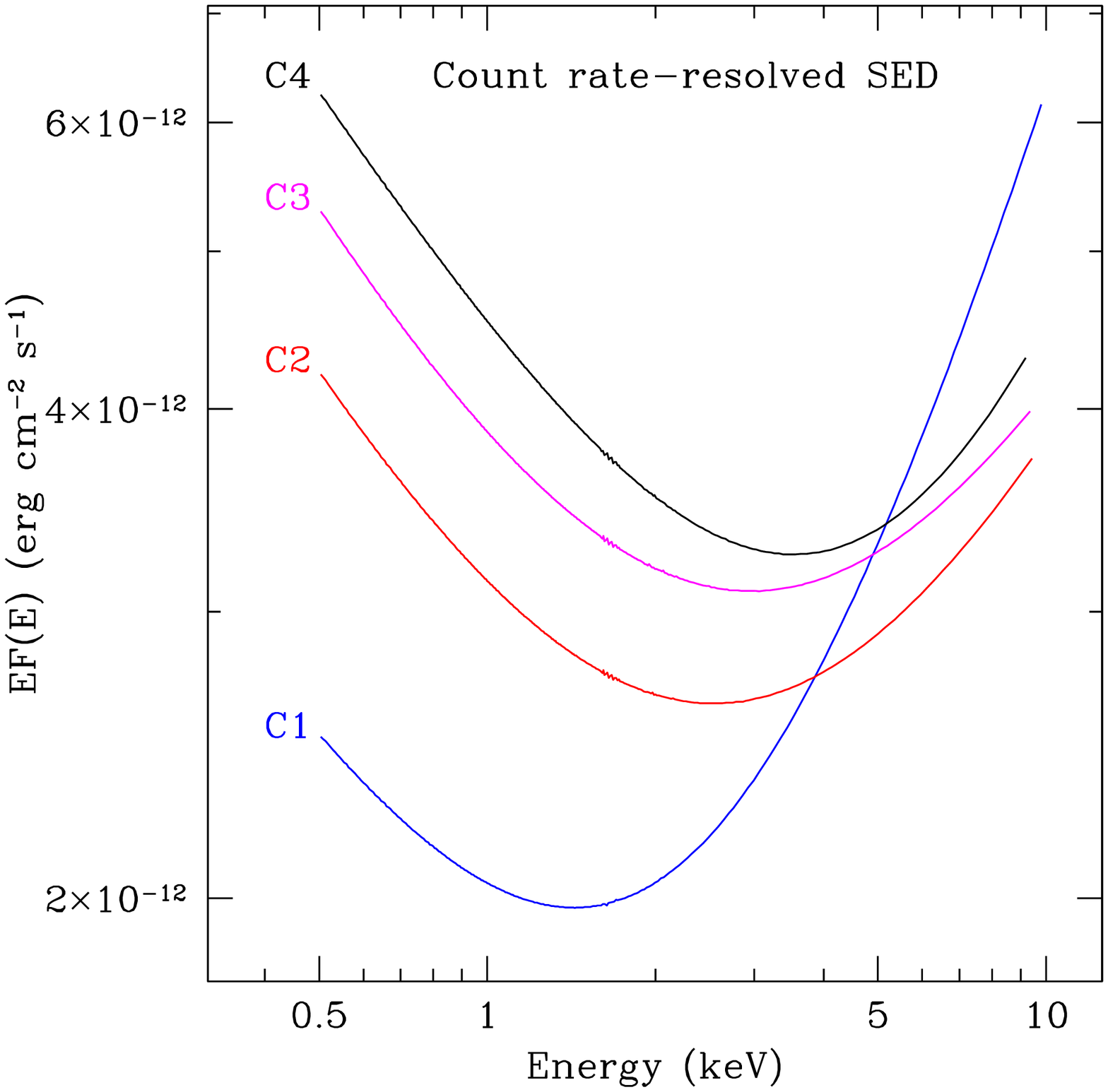} \caption { \footnotesize
The evolution of the unabsorbed X-ray SEDs (the sum of the
synchrotron and IC component) unfolded with the double power law.
With increasing total fluxes, the synchrotron emission extends to
higher energies, while the IC emission recedes from lower energies.
The T5 and C1 interval are identical, which are affected by the high
background.} \label{fig:xsed}
\end{figure}

Figure~\ref{fig:xsed} clearly shows the evolution of the 0.5--10~keV
X-ray SEDs unfolded with the double power law. Except for the T5 and
C1 SEDs, the SED slopes below the trough energies appear to not
change significantly with the fluxes, whereas the SED slopes above
the trough energies show larger changes with the fluxes. Due to
small changes of the IC fluxes, the shift of the SED troughs to
higher energies with increasing fluxes is primarily caused by the
increases of the synchrotron fluxes. The significant hardening of
the T5 and C1 SEDs above the trough energies might be affected by
the high background, since their SED slopes below the trough
energies are similar to those of other intervals.


\section{OM data analysis}\label{sec:om}

The OM observation provides optical/UV data simultaneous to the
X-ray observation. The data at each filter provide a time coverage
of $\sim 15$~ks, showing short-term variability of S5~0716+714 in
the optical-UV wavelengths.

\begin{figure} \epsscale{0.45}
\plotone{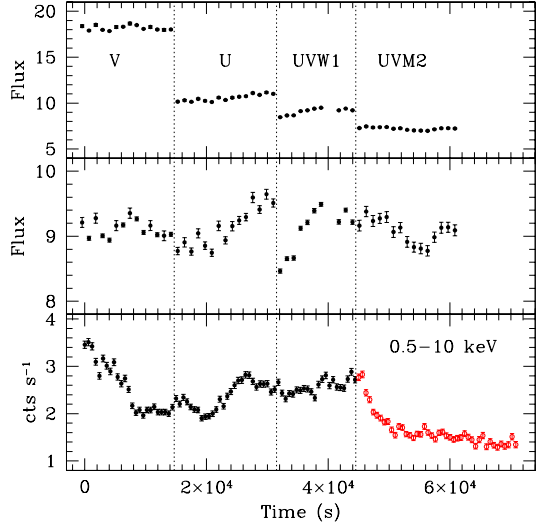} \caption {\footnotesize The upper panel plots the
light curves in different OM filters (the error bars on fluxes are
smaller than the symbol sizes). The middle panel shows the scaled
UVW1 band light curve. The fluxes are in unit of $10^{-26}~{\rm
ergs}~{\rm cm}^{-2} ~{\rm s}^{-1} ~{\rm Hz}^{-1}$. The bottom panel
presents the 0.5--10~keV light curve. The vertical dotted lines
separate the time intervals covered by different OM filters,
allowing an easy comparison between the light curves of different OM
filters and the X-ray light curve over the same time range. }
\label{fig:om}
\end{figure}

The count rates are converted into fluxes using the conversion
factors for a white dwarf, as recommended by the OM calibration
manual. Using the extinction curve of Cardelli et al. (1989) and the
extinction parameters of $R_V=3.1$ and $A_V=0.102$, we correct the
fluxes with the extinction coefficient at the effective wavelength
of each filter. The value of $A_V$ is taken from NED, which was
obtained by following Appendix B of Schlegel et al. (1998). The OM
light curves after extinction correction are shown in the top panel
of Figure~\ref{fig:om} with filter name indicated.

The fluxes gathered with different filters can not be used to do a
straightforward comparison with the X-ray light curve for the whole
observational length. We thus scale all the V, U and UVM2 fluxes to
the fluxes at the effective wavelength of the UVW1 filter ($\lambda
= 291$~nm). We assume a power law spectrum between any pair of
filters, whose spectral index is calculated with the average fluxes
at the two filters. The scaled UVW1 light curve is shown in the
middle panel of Figure~\ref{fig:om}. By adopting a constant spectral
index to scale the fluxes from one filter to UVW1, the scaled UVW1
light curve keep the same shape as the original one. For each
filter, our scaling law changes its flux level only. The 0.5--10~keV
light curve is also shown in the bottom panel of Figure~\ref{fig:om}
for a visual comparison with the scaled UVW1 light curve. It is
important to emphasize that the scaled UVW1 light curve is just used
to make a general comparison rather than to build strict correlation
with the X-ray light curve, since the scaled UVW1 fluxes might be
sufficiently different from the true UVW1 fluxes due to the known
spectral variability of the source in the optical-UV range.

The scaled UVW1 light curve does not closely track the X-ray one
throughout the observation. The most significant difference might be
the sharp transition from the last U-to-UVW1 scaled flux to the
first UVW1 flux, which does not have a correspondence in the X-ray
light curve. Nevertheless, it is certain that the variability
amplitude is significantly smaller in the UVW1 band than in the
X-ray band.

It is also possible to compare the OM light curves in different
filters with the X-ray light curve over the same time intervals,
respectively. It is clear that the V, UVW1 and UVM2 light curves do
not follow the X-ray light curve. The decay of $\sim 8$~ks long in
the X-ray light curve is not seen in the V band light curve.
Instead, the V light curve seems to show a well-defined micro-flare.
The UVW1 light curve displays a rise of $\sim 6$~ks long, which is
not clear in the corresponding X-ray light curve. The UVM2 light
curve shows a decay followed by a rise, whereas the X-ray light
curve exhibits a rapid decay followed by a slow decay. However, the
U light curve appears to track the X-ray light curve, where the
peaks in the U light curve show a delay of $\sim 2$~ks with respect
to the ones in the X-ray light curve.

\begin{figure} \epsscale{0.45}
\plotone{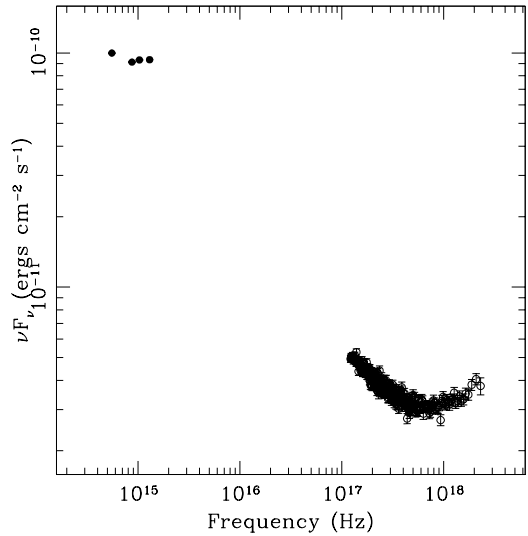} \caption {\footnotesize The average SED in the
optical-UV and X-ray wavelengths. The X-ray data are unfolded with
the double power law model. The error bars on the four optical-UV
fluxes are smaller than the symbol sizes. } \label{fig:omsed}
\end{figure}

The OM data also extend the synchrotron SED of the source to the
optical-UV range, where the synchrotron emission may peak around.
Figure~\ref{fig:omsed} plots the average optical-UV-X-ray SED. The
four optical-UV data points appear to be roughly on the
extrapolation of the soft X-ray SED, suggesting the same origin of
the optical-UV and soft X-ray emission.


\section{Discussion and Conclusions}\label{sec:disc}

We perform a detailed spectral and temporal analysis for the second
\xmm\ observation of S5~0716+714. Most of our results are in
agreement with previous results obtained from the first \xmm\
observation (FE06) and other X-ray observations (e.g., Giommi et al.
1999; Tagliaferri et al. 2003). Nevertheless, we also find some new
phenomena, adding new clues to better understand the underlying
physical processes taking place in the source.

The concave X-ray spectra of S5~0716+714 can be disentangled into
two power law components. The steep power law ($\Gamma \sim 2.6$)
component is interpreted as the high energy tail of the synchrotron
emission, whereas the flat power law ($\Gamma \sim1.2$) component is
ascribed to the low energy side of the IC emission. FE06 obtained
similar results with the first \xmm\ observation. It is worth noting
that the photon indices of the steep power law component are similar
to those of HBLs in the hard X-ray band (e.g., PKS~2155--304: Zhang
2008), supporting the interpretation of the steep power law
component as the synchrotron tail.

The X-ray variability amplitude of HBLs monotonically increases with
higher energy (e.g., PKS~2155--304: Zhang et al. 2005; Mrk~421:
Zhang et al. 2010), which is thought to be the signature of
synchrotron emission. However, LBLs are highly variable in the soft
X-rays, whereas they show little variability in the hard X-rays
(e.g., Giommi et al. 1999). Our results demonstrate that S5~0716+714
is indeed strongly variable in the soft X-rays, showing the maximum
variability by a factor of $\sim4$ throughout the whole observation
and several episodes of rapid variations on timescales of hours. In
a sharp-cut contrast, the hard X-ray fluxes of the source are much
less variable, exhibiting only $\sim50\%$ change between the minimum
and maximum count rates and no rapid events. For the first time, we
quantify the energy dependence of the variability amplitude for
S5~0716+714. The variability amplitude increases from the 0.3--0.5
to 0.5--7.5~keV band, but from the 0.75--1~keV band, it decreases
with higher energy. Moreover, the OM data suggest lower variability
amplitude in the UV band than in the soft X-ray band. The energy
dependence of the variability amplitude of S5~0716+714 is thus
clearly different from those of HBLs.

During the first \xmm\ observation, the \fvar\ of S5~0716+714
amounts to $40\pm3\%$ in the 0.5--0.75~keV and $27\pm1\%$ in the
3--10~keV (FE06). Though longer duration of the second \xmm\
observation, the \fvar\ is only $32.03\pm2.12\%$ and $7.62\pm1.5\%$
in the corresponding energy bands. It is important to understand
whether the synchrotron or IC emissions or both are responsible for
the significant variations of S5~0716+714 in different energy bands.
It is usually argued that the variable high energy tail of the
synchrotron emission is responsible for the highly variable soft
X-ray emission, whereas the "stable" low energy side of the IC
emission accounts for the lack of variability in the hard X-ray band
(e.g., Giommi et al. 1999; Tagliaferri et al. 2003). This argument
is somewhat ambiguous, since the two \xmm\ observations clearly
demonstrate that the hard X-ray fluxes are also highly variable,
although their variability amplitudes are significantly smaller than
the soft X-ray ones. The hard X-ray variations could be due to
either the IC or synchrotron variations. It is certain that the
synchrotron component is strongly variable on short timescales,
whereas it is unclear whether the IC component is variable on
similar timescales. The spectral analysis by FE06 showed that the IC
component might be variable on timescales of hours, which is,
however, not supported by our results. Although the IC component
contributes more to the total hard X-ray fluxes, we still assume
that the hard X-ray variations might be controlled by the
synchrotron tail. With increasing energies, the increasing dilutions
of the "stable" IC component to the increasing variations of the
synchrotron tail, might result in the observed overturn of the
energy-dependent variability amplitude for S5~0716+714.

The model-independent hardness-ratio analysis shows that the
0.5--10~keV spectra of S5~0716+714 soften when it brightens. The
phenomenon, also noticeable in the first \xmm\ observation (FE06;
Foschini et al. 2006), was already found in the \sax\ observations
of the source (Giommi et al. 1999). The softer-when-brighter
phenomenon is interpreted in terms of the high energy tail of the
synchrotron emission extending more to higher energies when the
source brightens (e.g., Giommi et al. 1999). Nevertheless, the soft
X-ray spectra appear to harden when it brightens, though the trend
is not significant. The variability property found in the soft X-ray
band of the source with ROSAT observations (Cappi et al. 1994) is
similar to what we found here. The harder-when-brighter trend is
analogous to those of HBLs, presenting a strong evidence that the
soft X-ray emission of S5~0716+714 is dominated by the synchrotron
tail.

Although the inter-band time lags and the related energy dependence
have been detected in a few X-ray bright HBLs, they have not be
firmly detected in LBLs yet. FE06 claimed that the lags of
$\gtrsim100$~s were not present in the first \xmm\ observation of
S5~0716+714. However, we found that the 0.3--0.5~keV variations lag
the 3--10~keV ones by $\sim 1000$~s in the second \xmm\ observation.
We also found a weak evidence that the lags increase with larger
energy differences. In at least one episode of the optical-UV light
curves, the U band variations might lag the X-ray variations by
$\sim 2000$~s. As far as we know, it is the first evidence for a
definite detection of soft lag in the X-ray variability of
S5~0716+714 and possibly LBLs. Interestingly, the soft lags and the
related energy dependence of S5~0716+714 are similar to what have
been detected in HBLs (e.g., Kataoka et al. 2000; Zhang et al.
2010). The similarity suggests that the X-ray variations of the
source have the same origin as those of HBLs, i.e., the variations
of the synchrotron tail. Therefore, the hard X-ray fluxes of the
source are dominated by the IC component, but the hard X-ray
variations might be still controlled by the synchrotron tail, as
already suggested by the energy dependent variability amplitude. If
we assume that both the soft and hard X-ray variations are caused by
the synchrotron tail, the observed soft lags provide a way to
constrain the physical parameters of the emitting region (e.g.,
Zhang \et 2002),
\begin{equation}
B\delta^{1/3} =209.91 \times \left (\frac{1+z}{E_{\rm
l}}\right)^{1/3}
    \left [\frac{1 - (E_{\rm l}/E_{\rm h})^{1/2}}
        {\tau_{\rm soft}} \right ]^{2/3} \quad {\rm G}  \,,
\label{eq:soft}
\end{equation}
where $\tau_{\rm soft}$ is the observed soft lag (in second) between
the low ($E_{\rm l}$) and high ($E_{\rm h}$) energy (in keV), $z$
the source's redshift, $B$ the magnetic field (in G) and $\delta$
the bulk Doppler factor of the emitting region. If adopting
$\tau_{\rm soft} \sim 1000$~s between the 0.3--0.5 and 3--10~keV,
one gets $B\delta^{1/3} \sim 2.56$~G. During a model fit to the SED
involving the first \xmm\ observation of S5~0716+714, Foschini et
al. (2006) assumed $B=3$~G and $\delta =16.7$, i.e., $B\delta^{1/3}
= 7.67$~G. A larger $B\delta^{1/3}$ implies a smaller $\tau_{\rm
soft}$, qualitatively consistent with the soft lags of no larger
than 100~s claimed by FE06.

The unprecedented high signal-to-noise ratio PN observation allows
us to disentangle the synchrotron and IC components in the X-ray
band of S5~0716+714 and to synchronously study the variations of the
two components on timescales of hours. The results, obtained with
the divisions of the observation by individual episodes and by count
rate levels, are consistent with each other. The synchrotron photon
indices are constrained in a limited range of $\Gamma \sim 2.5-2.7$,
which are typical of HBLs in the hard X-rays (e.g., PKS~2155--304:
Zhang 2008). The IC photon indices show relatively large changes
($\Gamma \sim 0.9-1.4$). Due to large uncertainties, the synchrotron
photon indices might be consistent with each other within the error
bars, which could be also true for the IC photon indices. Although
the synchrotron and IC photon indices do not correlate with the
total fluxes, the synchrotron spectra appear to harden with higher
synchrotron fluxes, and the IC spectra seem to soften with higher IC
fluxes. When the source brightens, the synchrotron fluxes increase,
while the IC fluxes decrease. The synchrotron fluxes also show
larger variations than the IC fluxes.

The X-ray SEDs convolved with the synchrotron and IC components
exhibit significant concave shapes. The crossing energies and the
SED trough energies increase with the increasing total fluxes. The
flux dependence of the crossing energies and the SED trough energies
is consistent with that of the first \xmm\ observation (FE06). We
further notice that the SED trough energies are smaller than the
crossing energies in most cases, indicating that they are not the
exact energies at which the synchrotron component transits to the IC
component in terms of the equal contributions of the two components
to the total fluxes. The SED evolution, characterized by the shifts
of the SED troughs to higher energies with higher fluxes, might be
mainly caused by the changes of the synchrotron normalization. The
changes of the synchrotron and IC photon indices and of the IC
normalization may affect the SED evolution in a weaker way.

The spectral and temporal behaviors of S5~0716+714 and its
synchrotron and IC emission are just the consequence of the peaks of
both the synchrotron and IC SEDs shifting to higher energies with
increasing fluxes. When the source brightens, the synchrotron peak
moves to higher energy. In turn, the high energy tail of the
synchrotron emission extends to higher energy. The synchrotron peak
is therefore more close to the observed X-ray band, bring on that
the synchrotron flux increases and the synchrotron spectrum hardens.
At the same time, the IC peak also shifts to higher energy,
incurring that the low energy end of the IC emission recedes from
lower energy to the observed X-ray band. The IC peak is thus more
far from the observed X-ray band.  As a result, the IC flux
decreases and the IC spectrum hardens. Accordingly, the synchrotron
flux anti-correlates with the IC flux when the source brightens. The
series of changes make the SED trough move to higher energy with
higher total flux. The high energy tail of the synchrotron emission
originates from the high energy electrons, showing strong and rapid
variations. The low energy end of the IC emission comes from the low
energy electrons, exhibiting small and slow variations. Soft lag is
expected if both the soft and hard X-ray variations are caused by
the cooling of the high energy electrons.

In conclusions, S5~0716+714 exhibits different X-ray variability
properties between the two \xmm\ observations. During the second
observation, it shows harder synchrotron and IC spectra and lower
variability amplitude. Even though the low energy end of the IC
emission contributes more to the hard X-ray fluxes than the high
energy tail of the synchrotron emission does, the energy dependence
of the variability amplitude suggests that the hard X-ray variations
might be dominated by the synchrotron tail. The large hard X-ray
variability amplitude suggests its synchrotron origin as well. More
importantly, soft lags are detected for the first time, also
supporting that the soft and hard X-ray variations are caused by the
same mechanism. When the the source brightens, the synchrotron
fluxes increase but the IC fluxes decrease. The synchrotron spectra
might harden with higher synchrotron fluxes, while the IC spectra
might soften with larger IC fluxes. The synchrotron tail shows
larger flux variations but smaller spectral variations than the IC
emission does. With higher fluxes, the crossing points between the
two components and the SED troughs move to higher energies. The
X-ray variability of S5~0716+714 is in accordance with the results
of its synchrotron and IC SED peaks shifting to higher energies with
higher fluxes. The decompositions of its X-ray emission into the
synchrotron and IC components are helpful to understand the
behaviours of both the low and high energy part of the electron
population.


\acknowledgments

This work is supported by the National Natural Science Foundation of
China (Project 10878011, 10733010 and 10473006) and by the National
Basic Research Program of China -- 973 Program 2009CB824800. This
research is based on observations obtained with \xmm, an ESA science
mission with instruments and contributions directly funded by ESA
Member States and NASA.


\clearpage
\begin{deluxetable}{llrccccccccc}
\tabletypesize{\scriptsize} \tablecolumns{11} \tabcolsep 3pt
\tablewidth{0pc} \tablecaption{Details of the \xmm\ EPIC-PN
observation for S5~0716+714}
\tablehead{ \colhead{} &\colhead{} &\colhead{Date} &\colhead{Time}
&\colhead{} &\colhead{} &\colhead{} &\colhead{Duration} &\colhead{On
Time} &\colhead{Live Time} &\colhead{} \\
\colhead{ObsID} &\colhead{Orbit} &\colhead{(UT)} &\colhead{(UT)}
&\colhead{Detector} &\colhead{Mode} &\colhead{Filter} &\colhead{(s)}
&\colhead{(s)} &\colhead{(s)} &\colhead{Count Rate\tablenotemark{a}}
}

\startdata
0502271401 &1427 &2007 Sep 24 &17:00:24--12:54:08\tablenotemark{b} &PN &SW\tablenotemark{c} &Thin &73917 &71564 &50125 &2.17 \\
\enddata
\tablenotetext{a}{Background subtracted mean count rate in the
0.5--10~keV band.} \tablenotetext{b}{Next day.}
\tablenotetext{c}{SW: imaging small window.} \label{tab:epic}
\end{deluxetable}

\begin{deluxetable}{lccccccc}
\tabletypesize{\scriptsize} \tablecolumns{8} \tabcolsep 3pt
\tablewidth{0pc}

\tablecaption{Results of \xmm\ OM observation (Imaging Mode)}

\tablehead{ \colhead{Filter} &\colhead{Wavelength (nm)}
&\colhead{Date (UT)} &\colhead{Exp. time (s)} &\colhead{Exp. number}
&\colhead{Count Rate\tablenotemark{a}}
&\colhead{Magnitude\tablenotemark{b}}
&\colhead{Flux\tablenotemark{c}} }

\startdata

V    &543 &16:51:25--21:07:47 &800 &14 &$67.03\pm0.10$  &$13.39\pm0.01$ &$1.836\pm0.003$ \\
U    &344 &21:13:08--01:48:11\tablenotemark{d} &800 &15 &$117.52\pm0.19$ &$13.08\pm0.01$ &$2.652\pm0.004$ \\
UVW1 &291 &01:53:31--05:25:07 &800 &10 &$56.44\pm0.06$  &$12.83\pm0.01$ &$3.202\pm0.003$ \\
UVM2 &231 &05:30:28--10:05:30 &800 &15 &$14.09\pm0.03$  &$12.90\pm0.01$ &$4.042\pm0.009$ \\

\enddata
\tablenotetext{a}{Background subtracted mean count rate.}
\tablenotetext{b}{Average instrumental magnitude.}

\tablenotetext{c}{Extinction corrected mean flux in unit of
$10^{-14}~{\rm ergs ~ cm^{-2} ~ s^{-1}~\AA^{-1} }$. }

\tablenotetext{d}{Next day.} \label{tab:om}
\end{deluxetable}


\begin{deluxetable}{lccccccc}
\tablecolumns{8} \tabletypesize{\scriptsize} \tabcolsep 5pt
\tablewidth{0pt} \tablecaption{Results of X-ray spectral fits to the
average X-ray spectrum\tablenotemark{a}}

\tablehead{ \colhead{ } &\colhead{ } &\colhead{$E_{\rm
break}$(keV)/} &\colhead{$\Gamma_2$/} &\colhead{ } &\colhead{ }
&\colhead{Flux\tablenotemark{d}}
&\colhead{Flux\tablenotemark{e}}  \\

\colhead{Model} &\colhead{$\Gamma_1$\tablenotemark{b}}
&\colhead{$\beta$\tablenotemark{c}} &\colhead{$kT$(keV)}
&\colhead{$\chi^2_{\nu}$/dof} &\colhead{Prob.}
&\colhead{(2--10~keV)} &\colhead{(0.5--10~keV)} }

\startdata

Single power law &$2.25\pm0.01$ &-- &-- &1.73/792 &$2.7\times10^{-33}$ &$4.39\pm0.06$ &$9.85\pm0.07$ \\

Broken power law &$2.38\pm0.02$ &$1.94^{+0.32}_{-0.11}$ &$1.92^{+0.03}_{-0.07}$ &1.05/790 &0.18 &$5.13\pm0.08$ &$10.54\pm0.08$ \\

Double power law &$2.56^{+0.09}_{-0.07}$ &-- &$1.22^{+0.19}_{-0.21}$ &1.00/790 &0.53 &$5.23\pm0.08$ &$10.66\pm0.09$ \\

Logarithmic parabola &$2.37\pm0.01$ &$-0.37\pm0.02$ &--  &1.01/791 &0.43 &$5.19\pm0.08$ &$10.62\pm0.09$ \\

Power law + black body &$2.40\pm0.02$ &-- &$2.24^{+0.13}_{-0.16}$ &1.02/790 &0.35 &$5.16\pm0.09$ &$10.59\pm0.09$ \\

Power law + bremss. &$1.93\pm0.03$ &-- &$0.38\pm0.03$ &1.04/790 &0.21 &$5.14\pm0.08$ &$10.55\pm0.08$ \\

\enddata
\tablenotetext{a}{The fits are performed in the 0.5--10~keV band.
The neutral hydrogen absorption column density is fixed to the
Galactic value. All quoted errors are $90\%$ confidence level
($\Delta\chi^2=2.706$) for one interesting parameter. }

\tablenotetext{b}{In the logarithmic parabola model, $\Gamma_1$ is
the photon index at 1~keV.}

\tablenotetext{c}{$\beta$ is the curvature parameter in the
logarithmic parabola model ($\beta<0$ means that the spectrum is
upward curved).}

\tablenotetext{d}{The unabsorbed 2--10~keV flux in unit of
$\rm{10^{-12}~ergs ~ cm^{-2}~ s^{-1}}$.}

\tablenotetext{e}{The unabsorbed 0.5--10~keV flux in unit of
$\rm{10^{-12}~ergs ~ cm^{-2}~ s^{-1}}$.}

\label{tab:spec}
\end{deluxetable}

\begin{deluxetable}{lccccc}
\tabletypesize{\scriptsize} \tablecolumns{6} \tabcolsep 3pt
\tablewidth{0pc} \tablecaption{ \fvar\ and lags\tablenotemark{a}}
\tablehead{ \colhead{Band (keV)} &\colhead{\fvar(\%)} &\colhead{$\rm
CCF_{\rm max}$} &\colhead{$\tau_{\rm peak}$(s)} &\colhead{$\tau_{\rm
cent}$(s)} &\colhead{$\tau_{\rm fit}$(s)} } \startdata
0.3--0.5  &$15.06\pm1.33$ &...  &...     &...     &... \\
0.5--0.75 &$17.25\pm1.51$ &0.91 &0       &$-379$  &$-210^{+188}_{-191}$ \\
0.75--1   &$16.54\pm1.52$ &0.90 &0       &$-394$  &$-364^{+190}_{-196}$ \\
1--3      &$13.41\pm1.20$ &0.90 &0       &$-554$  &$-593^{+184}_{-188}$ \\
3--10     &$7.508\pm1.43$ &0.59 &$-1200$ &$-908$  &$-1084^{+385}_{-392}$ \\
\enddata
\tablenotetext{a}{The \fvar\ and lags are estimated by excluding the
high background interval. All lags are measured with respect to the
0.3--0.5~keV band. A negative value indicates a soft lag.}
\label{tab:fvarlag}
\end{deluxetable}


\begin{deluxetable}{lcccccccc}
\tablecolumns{9} \tabletypesize{\scriptsize} \tabcolsep 5pt
\tablewidth{0pt} \tablecaption{Results of X-ray spectral fits with
the broken power law\tablenotemark{a}}

\tablehead{ \colhead{Interval} &\colhead{$\Gamma_1$}
&\colhead{$E_{\rm break}$} &\colhead{$\Gamma_2$}
&\colhead{K\tablenotemark{b}} &\colhead{$\chi^2_{\nu}$/dof}
&\colhead{Prob.} &\colhead{F$_{2-10~{\rm keV}}$\tablenotemark{c}}
&\colhead{F$_{0.5-10~{\rm keV}}$\tablenotemark{d}} }

\startdata
\multicolumn{9}{c}{Time resolved interval}\\
\cline{1-9}

T1 &$2.46^{+0.04}_{-0.03}$ &$2.22^{+0.17}_{-0.42}$ &$1.89^{+0.11}_{-0.06}$ &$2.79\pm0.03$ &1.02/400 &0.38 &$5.39\pm0.17$ &$11.68\pm0.18$ \\

T2 &$2.36^{+0.04}_{-0.03}$ &$1.84^{+0.18}_{-0.17}$ &$1.83\pm0.06$ &$1.99\pm0.03$ &1.08/398 &0.12 &$4.80\pm0.15$ &$9.27\pm0.16$ \\

T3 &$2.34^{+0.15}_{-0.02}$ &$2.77^{+0.33}_{-1.04}$ &$1.75^{+0.33}_{-0.11}$ &$2.45^{+0.02}_{-0.10}$ &1.00/448 &0.50 &$5.16\pm0.15$ &$10.66\pm0.17$ \\

T4 &$2.39\pm0.03$ &$2.00^{+0.29}_{-0.17}$ &$1.88^{+0.06}_{-0.08}$ &$2.39\pm0.02$ &1.05/517 &0.22 &$5.20^{+0.15}_{-0.14}$ &$10.58\pm0.15$ \\

T5 &$2.20^{+0.04}_{-0.05}$ &$1.81^{+0.31}_{-0.17}$ &$1.42^{+0.07}_{-0.13}$ &$1.31\pm0.02$ &0.99/668 &0.53 &$5.24^{+0.23}_{-0.21}$ &$8.16^{+0.24}_{-0.21}$ \\

\cline{1-9}
\multicolumn{9}{c}{Count rate resolved interval}\\
\cline{1-9}

C1 &$2.20^{+0.04}_{-0.05}$ &$1.81^{+0.31}_{-0.17}$ &$1.42^{+0.07}_{-0.13}$ &$1.31\pm0.02$ &0.99/668 &0.53 &$5.24^{+0.23}_{-0.21}$ &$8.16^{+0.24}_{-0.21}$ \\

C2 &$2.36\pm0.03$ &$1.88\pm0.18$ &$1.85\pm0.05$ &$1.99\pm0.02$ &1.07/535 &0.11 &$4.69\pm0.12$ &$9.16\pm0.13$ \\

C3 &$2.39\pm0.03$ &$1.88^{+0.31}_{-0.19}$ &$1.92^{+0.05}_{-0.07}$ &$2.45\pm0.02$ &0.94/513 &0.83 &$5.29\pm0.13$ &$10.81\pm0.14$\\

C4 &$2.42^{+0.04}_{-0.03}$ &$2.31^{+0.43}_{-0.46}$ &$1.90\pm0.12$ &$2.86\pm0.03$ &0.95/472 &0.78 &$5.57\pm0.15$ &$12.02\pm0.16$\\

\enddata
\tablenotetext{a}{See note a of Table~\ref{tab:spec}. }

\tablenotetext{b}{$K$ is the normalization factor in unit of
$\rm{10^{-3}~ keV^{-1} ~ cm^{-2}~ s^{-1}}$.}

\tablenotetext{c}{The unabsorbed 2--10~keV flux in unit of
$\rm{10^{-12}~ergs ~ cm^{-2}~ s^{-1}}$.}

\tablenotetext{d}{The unabsorbed 0.5--10~keV flux in unit of
$\rm{10^{-12}~ergs ~ cm^{-2}~ s^{-1}}$.}

\label{tab:fit:bknpl}
\end{deluxetable}


\begin{deluxetable}{lcccccccc}
\tablecolumns{9} \tabletypesize{\scriptsize} \tabcolsep 5pt
\tablewidth{0pt} \tablecaption{Results of X-ray spectral fits with
the double power law\tablenotemark{a}}

\tablehead{ &\colhead{$\Gamma_1$\tablenotemark{b}}
&\colhead{$\Gamma_2$\tablenotemark{c}} & & &\colhead{E$_{\rm
cross}$\tablenotemark{d}} &\colhead{Flux(Syn \%)\tablenotemark{e}}
&\colhead{Flux(Syn \%)\tablenotemark{f}} &\colhead{Flux(Syn \%)\tablenotemark{g}} \\

\colhead{Interval} &\colhead{(Syn.)} &\colhead{(IC)}
&\colhead{$\chi^2_{\nu}$/dof} &\colhead{Prob.} &\colhead{(keV)}
&\colhead{(0.5--2~keV)} &\colhead{(2--10~keV)}
&\colhead{(0.5--10~keV)}  }

\startdata
\multicolumn{9}{c}{Time resolved interval}\\
\cline{1-9}
T1 &$2.62^{+0.15}_{-0.10}$ &$1.18^{+0.34}_{-0.40}$ &1.03/400 &0.31 &$4.21\pm1.81$ &$6.29^{+0.07}_{-0.06}$(88.6\%) &$5.46\pm0.18$(47.1\%) &$11.75\pm0.19$(69.3\%) \\

T2 &$2.67^{+0.27}_{-0.17}$ &$1.36^{+0.27}_{-0.35}$ &1.07/398 &0.15 &$2.33\pm0.86$ &$4.50\pm0.05$(75.4\%) &$4.86\pm0.16$(29.8\%) &$9.36\pm0.17$(51.7\%) \\

T3 &$2.58^{+0.23}_{-0.13}$ &$1.27^{+0.36}_{-0.45}$ &0.95/448 &0.77 &$3.67\pm1.94$ &$5.48\pm0.05$(84.4\%) &$5.26\pm0.16$(43.1\%) &$10.73\pm0.17$(64.2\%) \\

T4 &$2.57^{+0.15}_{-0.09}$ &$1.14^{+0.31}_{-0.36}$ &1.03/517 &0.31 &$3.93\pm1.51$ &$5.39\pm0.05$(87.2\%) &$5.32\pm0.16$(44.3\%) &$10.71\pm0.17$(65.9\%) \\

T5 &$2.51^{+0.18}_{-0.13}$ &$0.84^{+0.22}_{-0.26}$ &0.96/668 &0.77 &$2.33\pm0.47$ &$2.94\pm0.03$(79.1\%) &$5.47^{+0.24}_{-0.23}$(23.1\%) &$8.42\pm0.24$(42.7\%) \\

\cline{1-9}
\multicolumn{9}{c}{Count rate resolved interval}\\
\cline{1-9}

C1 &$2.51^{+0.18}_{-0.13}$ &$0.84^{+0.22}_{-0.26}$ &0.96/668 &0.77 &$2.33\pm0.47$ &$2.94\pm0.03$(79.1\%) &$5.47^{+0.24}_{-0.23}$(23.1\%) &$8.42\pm0.24$(42.7\%) \\

C2 &$2.61^{+0.18}_{-0.12}$ &$1.29^{+0.24}_{-0.29}$ &1.07/535 &0.14 &$2.83\pm0.92$ &$4.49\pm0.04$(79.7\%) &$4.77\pm0.13$(35.1\%) &$9.26\pm0.14$(56.7\%) \\

C3 &$2.66^{+0.21}_{-0.13}$ &$1.39^{+0.25}_{-0.31}$ &0.90/513 &0.94 &$2.78\pm1.04$ &$5.54\pm0.05$(81.1\%) &$5.38\pm0.14$(38.3\%) &$10.92\pm0.15$(60.0\%) \\

C4 &$2.53^{+0.11}_{-0.07}$ &$0.93^{+0.39}_{-0.44}$ &0.93/472 &0.87 &$5.47\pm2.46$ &$6.45\pm0.06$(93.4\%) &$5.70\pm0.17$(55.8\%) &$12.15\pm0.18$(75.8\%) \\

\enddata
\tablenotetext{a}{See note a of Table~\ref{tab:spec}. }

\tablenotetext{b}{$\Gamma_1$ is the synchrotron (Syn.) photon
index.}

\tablenotetext{c}{$\Gamma_2$ is the IC photon index.}

\tablenotetext{d}{$E_{\rm cross}$ is the crossing energy at which
the synchrotron and the IC component contribute equally to the total
flux. The errors on $\ecross$ are propagated from the $1\sigma$
errors in the photon indices and the normalization (i.e., fluxes) of
the double power law. The $1\sigma$ errors in the photon indices and
the normalization are approximately obtained by averaging their
$1\sigma$ two-sided errors from the spectral fits. }

\tablenotetext{e}{The unabsorbed 0.5--2~keV flux in unit of
$\rm{10^{-12}~ergs ~ cm^{-2}~ s^{-1}}$. The number in the
parenthesis indicates the contribution of the synchrotron
component.}

\tablenotetext{f}{The unabsorbed 2--10~keV flux in unit of
$\rm{10^{-12}~ergs ~ cm^{-2}~ s^{-1}}$. The number in the
parenthesis indicates the contribution of the synchrotron
component.}

\tablenotetext{g}{The unabsorbed 0.5--10~keV flux in unit of
$\rm{10^{-12}~ergs ~ cm^{-2}~ s^{-1}}$. The number in the
parenthesis indicates the contribution of the synchrotron
component.}

\label{tab:fit:2pl}
\end{deluxetable}

\clearpage

\begin{figure}\epsscale{0.6}
\plotone{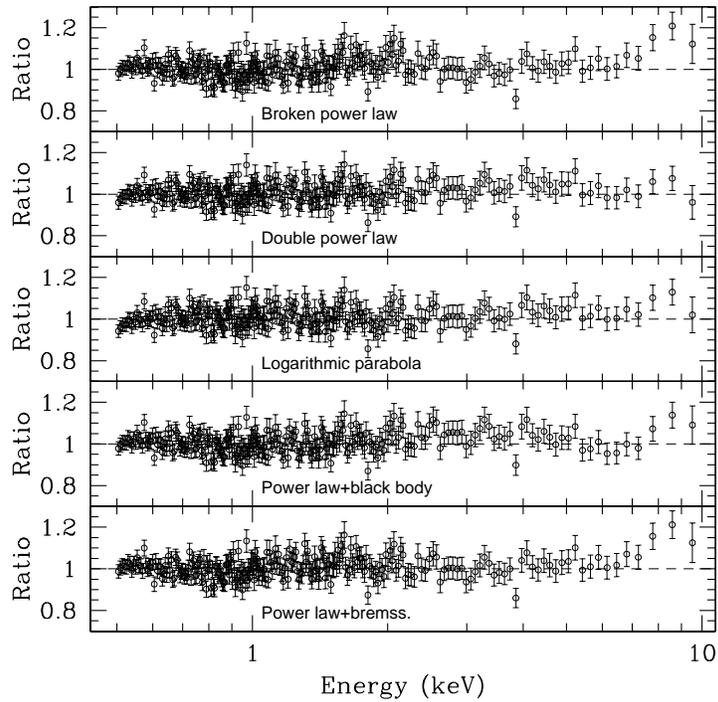} \caption {\footnotesize The data-to-model ratios
for the best fits with various models. The double power law model
presents the most significant fit. Although the fits are acceptable,
other models leave a visible hard tail above $\sim8$~keV. }
\label{fig:ratio} \label{fig:ratio}
\end{figure}

\end{document}